\documentclass[preprintnumbers, amssymb,amsmath,aps,prl,twocolumn,floatfix,nofootinbib,superscriptaddress,showpacs]{revtex4-1}

\usepackage{epsfig}
\usepackage{bm}
\usepackage{amssymb}
\usepackage{amsmath}
\usepackage[dvipsnames]{xcolor}
\usepackage[colorlinks,
  linkcolor=blue,
  anchorcolor=black,
  citecolor=blue
]{hyperref}
\usepackage{soul}
\usepackage[capitalise]{cleveref}
\usepackage[scr=boondoxo]{mathalfa}

\newcommand{\dd}{\mathrm{d}}
\newcommand{\abs}[1]{\left\vert#1\right\vert}

\newcommand{\Remy}{{\rm Re}}
\newcommand{\Immy}{{\rm Im}}

\newcommand{\bs}[1]{\boldsymbol{#1}}

\newcommand{\secprl}[1]{{\noindent\it {\color{blue}#1} }}

\makeatletter
\def\maketitle{
  \@author@finish
  \title@column\titleblock@produce
  \suppressfloats[t]
  \let\and\relax
  \let\affiliation\@gobble
  \let\@AAC@list\@empty
  \let\@AFF@list\@empty
  \let\@AFG@list\@empty
  \let\@AF@join\@AF@join@error
  \let\@address\@empty
  \let\thanks\@gobble
  \let\abstract\@undefined\let\endabstract\@undefined
  \titlepage@sw{
    \vfil
    \clearpage
  }{}
}
\makeatother

\begin{document}

\title{Nucleon Energy Correlators as a Probe of Light-Quark Dipole Operators at the Electron-Ion Collider}

\author{Yingsheng Huang}
\email{yingsheng.huang@outlook.com}
\affiliation{School of Physics, Central South University, Changsha 410083, China}
\affiliation{Department of Physics and Astronomy, University of Utah, Salt Lake City, UT 84112, USA}

\author{Xuan-Bo Tong}\email{xutong@jyu.fi}
\affiliation{Department of Physics, University of Jyväskylä, P.O. Box 35, 40014 University of Jyväskylä, Finland}
\affiliation{Helsinki Institute of Physics, P.O. Box 64, 00014 University of Helsinki, Finland}

\author{Hao-Lin Wang}\email{whaolin@m.scnu.edu.cn}
\affiliation{State Key Laboratory of Nuclear Physics and
  Technology, Institute of Quantum Matter, South China Normal
University, Guangzhou 510006, China}
\affiliation{Guangdong Basic Research Center of Excellence for
  Structure and Fundamental Interactions of Matter, Guangdong
  Provincial Key Laboratory of Nuclear Science, Guangzhou
510006, China}

\begin{abstract}
  We propose nucleon energy correlators (NECs) as a novel framework to probe electroweak light-quark dipole operators in deep inelastic scattering with an unpolarized nucleon. These operators encode chirality-flipping interactions, whose effects are usually quadratically suppressed in unpolarized cross sections. We construct a chiral-odd quark NEC that accesses quark transverse spin via azimuthal angle asymmetries in the energy flow of the target fragmentation region. These asymmetries serve as clean and powerful observables, enabling linear constraints on the quark dipole couplings. Unlike existing methods, our approach requires neither polarized nucleon beams nor final-state hadron identification, relying instead on fully inclusive calorimetric measurements. This work establishes one of the first applications of energy correlator observables to new physics searches and opens a promising direction for precision studies of chirality-flipping effects at electron-ion colliders.
\end{abstract}

\maketitle

\secprl{Introduction}~
The search for new physics (NP) beyond the standard model (SM) has been a major focus of collider experiments.
The lack of direct signals suggests that NP may emerge at an energy scale $\Lambda$ well beyond the current experimental reach. The standard model effective field theory (SMEFT) offers a systematic framework to describe such NP effects via higher-dimensional operators built from SM fields~\cite{Buchmuller:1985jz,Arzt:1994gp,Grzadkowski:2010es}. Among these, dimension-6 electroweak (EW) dipole operators involving light fermions are particularly intriguing for their chirality-flipping structure\footnote{See, e.g., the quark dipole operator $(\bar{\mathcal{Q}}\sigma^{\mu\nu}d) HB_{\mu\nu}$ in Eq.~\eqref{eq:SMEFTops}.}~\cite{Alonso:2013hga,Kley:2021yhn,Aebischer:2021uvt,Ardu:2025rqy,Boughezal:2021tih,Allwicher:2022gkm,daSilvaAlmeida:2019cbr,Li:2024iyj,Gauld:2024glt,Li:2025fom,Escribano:1993xr,Kopp:1994qv,Boughezal:2023ooo,Wen:2023xxc,Wen:2024cfu,Wen:2024nff}. Such effects are highly suppressed in the SM but essential for understanding anomalous magnetic moments {(AMM)} and $P$, $T$ violating electric dipole moments {(EDM)} of leptons and baryons ~\cite{Hecht:2001ry,Pitschmann:2014jxa,Liu:2017olr,
Fukuyama:2012np,Bhattacharya:2012bf,Chupp:2017rkp,Alarcon:2022ero,Keshavarzi:2022kpc,Gonzalez-Sprinberg:2024efj,Muong-2:2021ojo,Muong-2:2025xyk,Aliberti:2025beg,BESIII:2025vxm,ACME:2018yjb,Fan:2022eto,Abel:2020pzs}. However, constraints on these operators {at colliders} remain weak because the interference between dipole and SM amplitudes, scaling as ${\cal O}(\Lambda^{-2})$, typically vanishes in unpolarized observables~\cite{Escribano:1993xr,Kopp:1994qv,daSilvaAlmeida:2019cbr,Boughezal:2021tih,Allwicher:2022gkm,Li:2024iyj,Gauld:2024glt,Li:2025fom}. Consequently, the leading effects arise at ${\cal O}(\Lambda^{-4})$, resulting in reduced sensitivity and contamination from dimension-8 operators.

A new strategy to overcome this issue was recently proposed in~\cite{Boughezal:2023ooo,Wen:2023xxc}, which leverages the transverse spin of fermions. Since transverse spin states are superpositions of helicity eigenstates, they allow interferences between opposite-helicity components, restoring sensitivity to the dipole-SM interference at ${\cal O}(\Lambda^{-2})$. Building on this insight, several observables have been designed for the Electron-Ion Collider~(EIC) and future lepton colliders~\cite{Wen:2023xxc,Wen:2024cfu,Wang:2024zns,Boughezal:2023ooo,Wen:2024nff}. While transverse polarization of electron beams can be generated and controlled in accelerators, accessing transversely polarized quarks is more difficult due to color confinement. Existing approaches require non-perturbative correlations between quark spin and a transverse reference vector, either via nucleon polarization or via the transverse momenta of final-state hadrons, encoded in transversity parton distribution functions~(PDFs) and dihadron fragmentation functions~\cite{Boughezal:2023ooo,Wen:2024cfu}. Nevertheless, these methods either demand polarized nucleon beams, which compromise luminosity, or rely on semi-inclusive measurements that require particle identification and multi-particle tracking.

To address these limitations, we propose nucleon energy correlators (NECs) as an innovative approach to probe light-quark EW dipole operators in inclusive deep inelastic scattering (DIS) with an unpolarized nucleon. Recently introduced in~\cite{Liu:2022wop} as an extension of energy correlators from final-state jets to nucleon structure, NECs enable the imaging of partonic dynamics through correlations between initial-state partons and final-state energy flux in the target fragmentation region (TFR)~\cite{Liu:2022wop,Cao:2023oef,Cao:2023qat,Liu:2023aqb,Liu:2024kqt,Chen:2024bpj,Mantysaari:2025mht,Li:2023gkh,Guo:2024jch,Guo:2024vpe} (see \cite{Moult:2025nhu} for a review). By exploiting these correlations, we construct a chiral-odd quark NEC that accesses the quark transverse spin inside an unpolarized nucleon. This NEC ensures a nonzero interference between the dipole and SM amplitudes, generating unique azimuthal asymmetries in the energy distribution in DIS.
We demonstrate that these asymmetries exhibit clean and strong sensitivity to the quark dipole couplings. Crucially, our method works without polarized nucleon beams and requires only inclusive calorimetric measurements, making it an exceptional observable for probing chirality-flipping interactions at both existing and future electron-ion colliders.

\secprl{Quark transversity NEC }
Let us begin by considering quark NECs in an unpolarized nucleon, which moves rapidly along the $+\hat z$ direction with momentum $P$. The associated NECs are encoded in the following correlation function~\cite{Liu:2022wop,Chen:2024bpj}:
\begin{align}
  &{\cal M}^{[\Gamma]}(x,\theta,\phi) = \int \frac{\dd\eta^-}{4\pi} e^{-ix P^+ \eta^-}
  \notag \\
  &\qquad\times\langle P|\bar \psi(\eta^-) {\cal L}_n^{\dagger}(\eta^-)\Gamma{\cal E}(\theta,\phi) {\cal L}_n(0) \psi(0) |P\rangle~,
  \label{eq:NEEC_def}
\end{align}
which describes the conditional probability of finding a quark with the longitudinal momentum fraction $x$ inside the target nucleon, given the observation of an energy flux at the solid angle $(\theta,\phi)$
in the TFR. $\theta$ and $\phi$ denote the polar and azimuthal angles, respectively. Formally, the quark NECs can be viewed as the collinear PDFs modified by the insertion of the energy flow operator ${\cal E}(\theta,\phi)$, which is defined through its action on hadronic states~\cite{Sveshnikov:1995vi,Bauer:2008dt}
\begin{align}
  {\cal E}(\theta,\phi)|X \rangle=\sum_{i \in X} \frac{E_i}{E_N}\delta(\theta_i^2-\theta^2)\delta(\phi_i-\phi)|X \rangle~,
  \label{eq:op}
\end{align}
where $E_i$ denotes the energy of hadron $i$, normalized by the nucleon energy $E_N$. Here, we use the light-cone coordinates for a vector $a^\mu = (a^+,a^-,  \vec a_\perp) $ with $a^{\pm}=(a^0\pm a^3)/\sqrt{2}$ , and  ${\cal L}_n$ is the light-cone gauge link.

The correlation function
${\cal M}^{[\Gamma]}$ depends on the choice of the Dirac matrix $\Gamma$, which selects specific quark spin components.
For instance, the leading-twist unpolarized quark NEC is obtained with $\Gamma=\gamma^+$:
\begin{align}
  f_1^q(x,\theta^2)={\cal M}^{[\gamma^+]}(x,\theta,\phi)~,
  \label{eq:f1}
\end{align}
which is chiral-even and rotationally invariant in the transverse plane~\cite{Liu:2022wop}.

While prior studies have primarily focused on the chiral-even NECs~\cite{Liu:2022wop,Chen:2024bpj,Liu:2024kqt,Cao:2023oef,Cao:2023qat,Liu:2023aqb,Mantysaari:2025mht} (see a systematic classification in  \cite{Chen:2024bpj}), we extend the formalism to the chiral-odd components. In particular, the quark transversity NEC can be obtained by using the projector $\Gamma^\alpha=i \sigma^{\alpha+} \gamma_5$ with $\alpha=1,2$, analogous to that used for transversity PDFs. {The interpretation of quark transverse spin becomes explicit by rewriting $   \Gamma^\alpha=\gamma^+ ({\cal Q}_{+}^\alpha-{\cal Q}_{-}^\alpha)
$, where ${\cal Q}_{\pm}^\alpha=\frac{1}{2}\big(1\mp \gamma^5 \gamma^\alpha_\perp\big)$ projects onto quarks with transverse spin aligned or anti-aligned with the $\alpha_{\perp}$-direction~\cite{Jaffe:1996zw}.}
However, unlike transversity PDFs, which requires a transversely polarized nucleon~\cite{Jaffe:1991ra}, NECs can access the quark transverse spin even in an {\it unpolarized} nucleon, owing to the azimuthal dependence of the observed energy flux.

To explicitly illustrate this feature, we introduce a {unit} vector {$ \hat n_T^\mu=(0,0 ,\cos\phi, \sin\phi)$} to specify the transverse direction of the observed energy flux. {By rotational covariance and parity property,} the quark transversity NEC $h_1^{t,q}(x, \theta^2)$ is {parametrized} as
\begin{align}
  \epsilon_\perp^{\alpha \rho}   {\hat n_{T,\rho}}h_1^{t,q}(x, \theta^2)={\cal M}^{[i \sigma^{\alpha+}\gamma_5 ]}(x,\theta,\phi)~,
  \label{eq:h1}
\end{align}
where
$\varepsilon_\perp^{\mu\nu} = \varepsilon^{\mu\nu-+} $ with $\varepsilon^{0123}=1$. This equation shows that $h_1^{t}(x, \theta^2)$ captures azimuthal correlations between the quark's transverse polarization and the energy flow direction. {Physically, it describes the difference in probability of finding a quark polarized in the two opposite directions
perpendicular to  the transverse direction of the energy flow.} Being chiral-odd, the transversity NEC is off-diagonal in the helicity basis and naturally induces interference between amplitudes of opposite quark helicities. This provides a novel mechanism for generating chirality-flipping effects, enabling probes of dipole operators in unpolarized nucleon scatterings.

\begin{figure}[t!]
  \centering
  \includegraphics[width=0.45\textwidth]{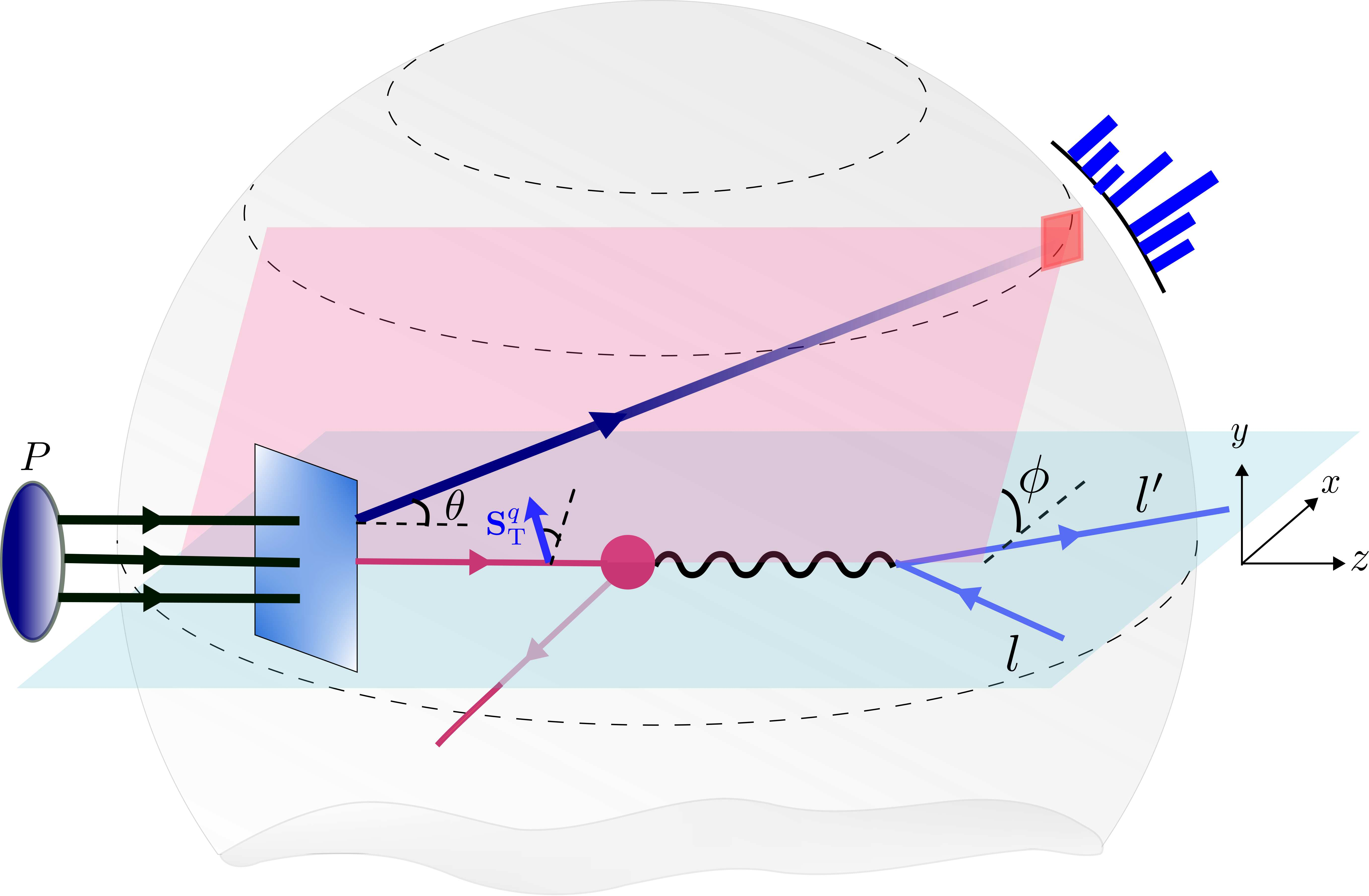}
  \caption{Illustration of the NEC mechanism for probing the SMEFT quark dipole operator in DIS. The blue transverse plate highlights the role of the transversity NEC, which selects the quark transverse spin $S_T^q$ in an unpolarized nucleon by measuring an energy flux from the target remnants. The magenta blob denotes the dipole interaction mediated by $\gamma$ or $Z$. The interference with the SM amplitude is implied.
  }

  \label{fig:NEC}
\end{figure}
\secprl{Probing the quark dipole operators with transversity NEC}
To demonstrate how the transversity NEC probes the SMEFT dipole couplings, we consider the energy pattern within the TFR in inclusive DIS with an unpolarized nucleon, $l +p\rightarrow l'+X$. Experimentally, this observable is determined from the angular distribution of the total energy deposited in the calorimeters. Following Refs.~\cite{Liu:2022wop,Chen:2024bpj}, we define the energy pattern cross section as:
\begin{align}
  \Sigma(\theta,\phi)=\sum_{i \in X} \int \dd \sigma^{l +p\rightarrow l'+X} \frac{E_i}{E_N}\delta(\theta^2-\theta_i^2) \delta(\phi-\phi_i)~.
  \label{eq:energy pattern}
\end{align}
We work in the Breit frame, where the momentum transfer ($q=l'-l$) is aligned along the $-\hat z$ direction, and the nucleon moves in the opposite direction. As shown in Fig.~\ref{fig:NEC}, the polar angle $\theta$ is measured with respect to the nucleon beam, and the azimuthal angle $\phi$ is defined relative to the lepton scattering plane. The standard DIS kinematic variables are also used: the photon virtuality $Q^2=-q^2$, the Bjorken variable $x_B=Q^2/(2P\cdot q)$, and the inelasticity $y=Q^2 /\left(s x_{\mathrm{B}}\right)$ with $\sqrt{s}$ denoting the center-of-mass energy of the $ep$ collision.

Let us first consider the case of an unpolarized electron beam. We are particularly interested in the following azimuthal modulations of the energy pattern:
\begin{align}
  \Sigma(\theta,\phi)=   \Sigma_{UU}(\theta)+   \Sigma^{\sin \phi}_{UU}(\theta)\sin \phi +  \Sigma^{\cos \phi}_{UU}(\theta)\cos \phi.
  \label{eq:SF}
\end{align}
We focus on the TFR, characterized by small polar angles ($\theta  P^+ \ll Q$), where the energy flux originates from the fragmentation of target remnants. Following the general arguments in \cite{Cao:2023oef,Chen:2024bpj}, the energy pattern cross section in this region factorizes into hard partonic scattering coefficients and the non-perturbative NECs, defined by Eq.~(\ref{eq:NEEC_def}). Because NECs encapsulate all the information regarding the energy flow, including its azimuthal $\phi$ dependence, the $\phi$-independent cross section $\Sigma_{UU}$ factorizes with unpolarized NECs, such as $f_1$ given in Eq.~\eqref{eq:f1}. By contrast, the $\phi$-dependent terms $\Sigma^{\sin \phi}_{UU}$ and $\Sigma^{\cos \phi}_{UU}$ are expected to factorize with the transversity NEC $h_1^t$, which encodes azimuthal correlations as seen in Eq.~(\ref{eq:h1}). Remarkably, after factorization, the hard scatterings remain the same as in standard inclusive DIS process~\cite{Liu:2022wop,Cao:2023oef,Chen:2024bpj}.

Within the SM, the energy pattern is isotropic in the transverse plane. Since quark helicity flip in DIS hard scattering is suppressed by light quark masses, only the chiral-even unpolarized quark NEC $f_1(x,\theta^2)$ contributes to $\Sigma(\theta,\phi)$ in the massless limit, while the chiral-odd contributions from the transversity NEC $h_1^{t}(x, \theta^2)$ vanish to all orders in $\alpha_s$. The only non-zero contribution arises from the azimuthally independent cross section $\Sigma_{UU}$.
At leading order in $\alpha_s$, the contribution from the photon exchanges takes the following form~\cite{Liu:2022wop}:
\begin{align}
  &\frac{ \dd \Sigma_{UU}}{\dd x_B \dd Q^2   }
  =\frac{2\pi\alpha^2_{\text{em}}}{ Q^4} ( y^2-2y+2)\sum_{q} Q_q^2 f_1^q(x_B,\theta^2)~,
  \label{eq:UU_SM}
\end{align}
where $\alpha_{\text{em}}=e^2/(4\pi)$ is the fine structure constant and $Q_q$ is the quark electric charge.

Although the $\sin \phi$ and $\cos \phi$ azimuthal modulations are absent within the SM, they could arise from chirality-flipping NP effects, such as the following dimension-6 SMEFT dipole interactions~\cite{Grzadkowski:2010es}:
\begin{align}
  \label{eq:SMEFTops}
  \mathcal{L}=&\frac{1}{\Lambda^2}\Big[C_{uW}(\bar{\mathcal{Q}}  \sigma^{\mu\nu} u) \tau^I \tilde{H} W^I_{\mu\nu}+C_{uB}(\bar{\mathcal{Q}}  \sigma^{\mu\nu} u)\tilde{H} B_{\mu\nu}
    \notag\\ +& C_{dW}(\bar{\mathcal{Q}}  \sigma^{\mu\nu} d) \tau^I  H W^I_{\mu\nu}+
  C_{dB}(\bar{\mathcal{Q}}  \sigma^{\mu\nu} d)  H B_{\mu\nu}\Big]+\textrm{h.c}.~,
\end{align}
where $\tau^I$ are the Pauli matrices, and $\tilde{H}_I=\epsilon_{IJ} H_J^*$. The field strength tensors $W_{\mu\nu}^I$ and $B_{\mu\nu}$ correspond to the $SU(2)_L$ and $U(1)_Y$ gauge fields, respectively. Here, $\mathcal{Q}$ denotes the left-handed quark doublet, while $u$ and $d$ represent the right-handed up- and down-type quark singlets. In this work, we restrict our attention to operators involving only the first generation and define their dipole couplings to photon and $Z$-boson $c_i$ as
\begin{align}
  \begin{aligned}
    c_{q\gamma} &= (v/{\sqrt{2}\Lambda^2})\left( c_W C_{qB} \pm s_W C_{qW} \right)~, \\
    c_{qZ} &= (v/{\sqrt{2}\Lambda^2})\left(-s_W C_{qB} \pm c_W C_{qW}  \right)~.
  \end{aligned}
  \label{eq:LEFT}
\end{align}
The plus (minus) sign is for the up (down) quark, $c_W=\cos \theta_W$ and $s_W=\sin \theta_W$ with $\theta_W$ being the Weinberg angle, and $v$ is the Higgs vacuum expectation value.

As illustrated in Fig.~\ref{fig:NEC}, these azimuthal modulations originate from the interplay between the SMEFT dipole operator and the quark transversity NEC $h_1^{t}(x, \theta^2)$. Being inherently chiral-odd, the dipole interaction induces a quark helicity flip during the hard scattering with the virtual photon or $Z$-boson. As a result, the dominant contributions emerge from the interference between the helicity-conserving SM amplitude and the helicity-flipping SMEFT amplitude at ${\cal O}(\Lambda^{-2})$.
Such interference necessarily couples to a chiral-odd quark NEC, namely the transversity NEC. As revealed in Eq.~(\ref{eq:h1}), the quark transverse spin is accessed through intrinsic azimuthal correlations with the energy flux direction $\vec{n}_T$, thereby yielding distinct $\sin \phi$ and $\cos \phi$ modulations.
Furthermore, these asymmetries are uniquely generated by the dipole operators, as other dimension-6 SMEFT operators either require transverse electron polarization or are suppressed by the electron Yukawa coupling.

We find that the $\sin \phi$-modulation is linearly determined by the real parts of the quark dipole couplings $c_{q\gamma}$ and $c_{qZ}$ through the interference between photon and $Z$-boson exchanges. The corresponding cross section is given by:
\begin{align}
  \label{eq:UUsin}
  &\frac{ \dd \Sigma_{UU}^{\sin \phi}}{\dd x_B \dd Q^2}
  =
  \frac{4\pi\alpha^2_{\text{em}}}{ ec_W s_W } \frac{y\sqrt{1-y} }{Q(Q^2+m_Z^2)}
  \sum_q h_1^{t,q}(x_B, \theta^2)
  \\ &\;\;\times\bigg\{\Big[\frac{2-y}{y} g_A^{q}g_V^e+ g_V^{q}g_A^e\Big]\frac{\Remy[c_{q\gamma}]}{c_W s_W}-Q_qg_A^e \Remy[c_{qZ}]\bigg \}~, \notag
\end{align}
where $g_V^f=T^3_f/2-Q_f s_W^2$ and $g_A^f= T^3_f/2$ are the vector and axial-vector couplings with $T^3$ as the third component of weak isospin. Similarly, the $\cos \phi$-modulation is sensitive to the imaginary parts of the dipole couplings and can be obtained from Eq.~\eqref{eq:UUsin} by replacing $\text{Re}[c_{q\gamma,Z}]$ with $-\text{Im}[c_{q\gamma,Z}]$. The anti-quark components in Eq.~\eqref{eq:UU_SM} and  Eq.~\eqref{eq:UUsin} can be obtained by taking $g_A^q\to -g_A^q$. Considering the EIC kinematics, we have neglected pure $Z$-boson exchanges.

Our analysis can be readily generalized to include a longitudinally polarized electron beam, yielding the cross section $\Sigma_{LU}^{\sin \phi (\cos \phi )}$ factorized in terms of the transversity NEC $h_1^{t}$. Owing to the parity violation induced by electron polarization, the azimuthal modulations can be generated via $\gamma\gamma$ interference, without $Q^2/m_Z^2$ suppression present in the unpolarized case $\Sigma_{UU}^{\sin \phi (\cos \phi )}$. As indicated later, this leads to an enhanced sensitivity to the dipole couplings. Complete analytical results are provided in the Supplemental Material~\cite{appendix}.
\nocite{Collins:1997sr,Kang:2012em,Artru:1989zv,Vogelsang:1997ak,Lu:2009ip,H1:2015ubc,Diaconu:2010zz,Christova:2020ahe,Zhang:2008ez}

\secprl{Calorimetric asymmetries and the non-perturbative input}
To quantify the effects of the SMEFT dipole operators, we introduce azimuthal angle asymmetries of the energy pattern. For an unpolarized electron beam, these are purely calorimetric asymmetries defined with respect to the lepton scattering plane:
\newcommand{\uu}{\mathscr u}
\begin{align}
  A^{\uu}_{UU} & = \frac{\pi}{2}{ \Sigma(\uu>0) - \Sigma(\uu<0) \over \Sigma(\uu>0) + \Sigma(\uu<0) }~,
  \label{eq:asymmetry}
\end{align}
where $\uu$ is taken to be either $\sin\phi$ or $\cos\phi$, isolating the real or imaginary part of the dipole couplings, respectively. $\Sigma(\uu>0) $ represents the energy pattern cross section integrated over the region with $\uu>0$:
\begin{align}
  \Sigma(\uu>0)\equiv\int ^{\theta_{\text{max}}^2}_{\theta_{\text{min}}^2}
  \dd \theta^2  {w}(\theta)\int^{2\pi}_0 \dd \phi\Sigma (\theta,\phi)\Theta(\uu)~,
  \label{eq:weighting}
\end{align}
and similarly for $\uu<0$. The polar-angle weight function $w(\theta)$ is introduced for later use. For a longitudinally polarized electron beam, one can analogously define the asymmetries $A_{L U}^{\uu}$ by combining the calorimetric asymmetry with the longitudinal spin asymmetry, as we specify in~\cite{appendix}.

Numerical evaluation of these asymmetries requires non-perturbative input for the quark NECs $f_1$ and $h_1^{t}$. While the unpolarized quark NEC $f_1$ can be extracted directly from $\Sigma_{UU}$ in the TFR~\cite{Liu:2022wop}, one can also measure the transversity NEC $h_1^{t}$ without assuming SMEFT interactions by tagging an additional hadron in the current fragmentation region. Such measurements are feasible at facilities including HERA, JLab, and the EIC.

Although direct experimental determinations of NECs are not yet available, a remarkable correspondence has been established between $\theta$-weighted moments of the NECs and $k_\perp$-weighted moments of transverse-momentum-dependent parton distributions~(TMDs), including the unpolarized quark NEC~\cite{Liu:2024kqt}. This correspondence can be straightforwardly extended to the transversity NEC. Specifically, we have
\begin{align}
  E_N\int \dd\theta^2\abs{\sin\theta} f_{1}(x,\theta^2)&=\int \frac{\dd^2 \bs k_\perp}{2\pi} |\bs k_\perp|\mathscr{f}_{1}(x,\bs k_\perp^2)~, \notag \\
  E_N\int \dd\theta^2  \abs{\sin \theta}  h_1^{t}(x, \theta^2)&=\int \frac{\dd^2 \bs k_\perp}{2\pi}  \frac{\bs k_{\perp}^2}{M}  \mathscr{h}_1^{\perp}(x, \bs k_\perp^2)~,
  \label{eq:TMDrelations}
\end{align}
where $\mathscr{f}_1$ and $\mathscr{h}_1^{\perp}$ are the unpolarized and Boer-Mulder quark TMDs~\cite{Boussarie:2023izj}, respectively. These relations allow us to infer the NECs from existing global fits of TMDs. In our analysis, we adopt the parameterization from~\cite{Barone:2009hw}, which assumes a Gaussian form for the $k_\perp$-dependence.  {The evolution effects are accounted for as described in~\cite{appendix}.}

To utilize the above relations, in Eq.~\eqref{eq:weighting} we choose the $\theta$-weight as $w(\theta)=|\sin \theta|$. The limits of the $\theta$-integration are determined by both theoretical and experimental considerations. The upper bound $\theta_{\text{max}}$ is imposed to ensure the validity of the factorization in the TFR~\cite{Cao:2023oef,Chen:2024bpj}. This corresponds to a transverse momentum cutoff $k_{\perp}^\text{max}\sim Q\theta_{\text{max}} $ in the $k_\perp$-integrals in Eq.~(\ref{eq:TMDrelations}). Due to rapid falloff of the Gaussian TMDs at large $k_\perp$,
this cutoff has negligible numerical impact on the integral. The lower bound $\theta_{\text{min}}$ is determined by the coverage of the calorimeter in the nucleon forward region. It imposes a minimal transverse-momentum threshold given by $k_{\perp}^{\text{min}}\approx E_{\text{min}} \theta_{\text{min}}$, where $E_{\text{min}}$ characterizes the energy resolution of calorimeters~\cite{AbdulKhalek:2021gbh}. Practically, we adopt $k_{\perp}^\text{min}\approx0~\text{GeV}$ and $k_{\perp}^\text{max}=5~\text{GeV}$. With these inputs, we have evaluated numerically both $A_{UU}^{\sin \phi}$ and $A_{L U}^{\sin \phi}$ for the EIC kinematics, and the detailed results can be found in the Supplemental Material~\cite{appendix}.

\begin{figure}
  \begin{centering}
    \includegraphics[width=\linewidth]{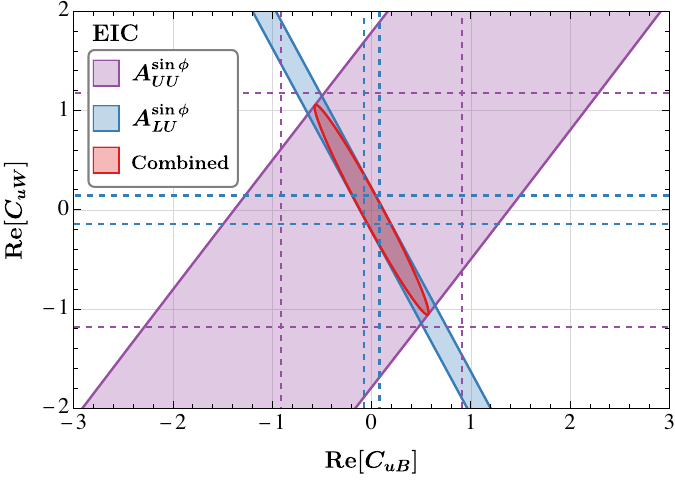}
    \caption{Projected $68\,\%$ C.L. constraints at the EIC on the two Wilson coefficients $\Remy[C_{uB}]$ and $\Remy[C_{uW}]$, assuming $\Lambda=1$ TeV. The limits are derived from the azimuthal asymmetries $A^{\sin\phi}_{UU}$ (purple), $A^{\sin\phi}_{LU}$ (blue), and their combination (red). The shaded regions represent the simultaneous two-coefficient constraints, while dashed lines show the limits for a single coefficient.
    }
    \label{fig:cons-CBW}
  \end{centering}
\end{figure}

\secprl{Projected sensitivity at the EIC}
To estimate the sensitivity to the dipole operators at the EIC, we perform a $\chi^2$ analysis based on the predicted asymmetries. We select a representative EIC configuration~\cite{AbdulKhalek:2021gbh} with a center-of-mass energy of $\sqrt{s}=105\ \text{GeV}$, which corresponds to the optimal integrated luminosity of $\mathcal{L} = 100\,\text{fb}^{-1}$. The analysis is carried out in $Q$ bins for $Q\in[10,\,60]$ GeV and $x\in[0.01,\,0.5]$, incorporating an inelasticity cut $0.1\leq y\leq 0.9$. We assume the experimental value consistent with the leading-twist SM prediction, and is negligible. With this setup, the statistical uncertainty $\delta A_i \simeq \pi/(2\sqrt{\mathcal{E}_\textrm{total}})$ is dominant over the systematic uncertainties. Here, $\mathcal{E}_\textrm{total}=(\Sigma(\uu>0)+\Sigma(\uu<0))\times \mathcal{L}$ is the accumulated energy deposition.

\cref{fig:cons-CBW} presents the projected constraints on the real parts of dimensionless Wilson coefficients $C_{uW}$ and $C_{uB}$ at $68\%$ C.L. using $\sin \phi$-asymmetry. Assuming single-operator dominance, $A_{UU}^{\sin \phi}$ constrain the Wilson coefficients almost at the $\mathcal{O}(1)$ level, as shown by the purple dashed lines. However, when both coefficients are considered simultaneously, the resulting constraint ellipse is elongated into a purple band, indicating a strong correlation between $C_{uW}$ and $C_{uB}$.

We observe that this correlation is significantly alleviated by incorporating the  $A_{LU}^{\sin\phi}$ asymmetry. For the EIC, we assume $70\%$ polarization of the electron beam~\cite{AbdulKhalek:2021gbh}. Under this setup, single-operator constraints can reach the $\mathcal{O}(0.01)\sim\mathcal{O}(0.1)$ level. Importantly, while $A_{LU}^{\sin\phi}$ retains strong correlations between coefficients, the orientation of the corresponding blue band is different from that of $A_{UU}^{\sin\phi}$. Their combination thus greatly improves the parameter space resolution, confining the allowed region to a narrow area highlighted in red.
Similar improvements are also observed in the constraints on $C_{dW}$-$C_{dB}$ and $c_{q\gamma}$-$c_{qZ}$ (see Supplemental Material~\cite{appendix}). The constraints on the imaginary parts can be obtained analogously from the $\cos \phi$-asymmetry, yielding identical sensitivities. {The constraints are stable against different nonperturbative inputs of NEC, as we analyze in~\cite{appendix}.} These results underscore the strong capability of the proposed asymmetries to unveil light-quark dipole interactions at the EIC.

\begin{figure}[t]
  \centering
  \includegraphics[width=\linewidth]{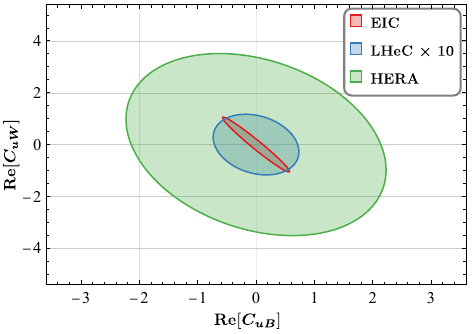}
  \caption{Comparison of projected $68\,\%$ C.L. constraints on $\text{Re}[C_{uB}]$ and $\text{Re}[C_{uW}]$ from HERA (green), the EIC (red), and the LHeC (blue), assuming $\Lambda=1$ TeV. The LHeC limits are scaled by a factor of $10$ for visibility. A detailed description of our analysis setup is provided in~\cite{appendix}.}
  \label{fig:3coll}
\end{figure}

Besides the future EIC, the absence of a polarized nucleon beam requirement for these asymmetries also opens up opportunities to use available data from HERA.  In Fig.~\ref{fig:3coll}, we compare the projected sensitivity from the combined-asymmetry analysis among the EIC, HERA~\cite{Klein:2008di}, and the LHeC~\cite{LHeC:2020van}. Although HERA benefits from higher-$Q^2$ coverage, its lower luminosity results in slightly weaker sensitivity compared to the EIC. However, it provides better discrimination between operators. With even higher energy and better luminosity, the LHeC could outperform both HERA and EIC by an order of magnitude.

Finally, let us comment on the photon-quark dipole coupling $c_{q\gamma}$ (given in Eq.~(\ref{eq:LEFT})), which can be constrained by measuring hadronic magnetic and electric dipole moments at low energy. While high-precision neutron EDM measurements yield strong bounds on $\text{Im}[c_{q\gamma}]$\cite{Kley:2021yhn,Kumar:2024yuu}, existing hadronic AMM measurements do not well constrain $\text{Re}[c_{q\gamma}]$, and thus $\text{Re}[C_{qW}]$ and $\text{Re}[C_{qB}]$. These real parts encode the $P$- and $T$-conserving NP effects and may produce sizable signatures without conflicting with neutron-EDM bounds. As shown in Fig.~\ref{fig:3coll}, our proposal provides sensitive limits on the real components; for explicit bounds on $\text{Re}[c_{q\gamma}]$, we refer the reader to the Supplemental Material~\cite{appendix}.
  Consequently, this letter is dedicated to searching for these $P$- and $T$-conserving NP effects at $ep$ colliders.

\secprl{Conclusion and discussion} In summary, we have shown that nucleon energy correlators provide a powerful framework to probe EW light-quark dipole operators in inclusive DIS with an unpolarized nucleon. By exploiting azimuthal correlations of the energy flux in the TFR, we introduce a chiral-odd NEC sensitive to the transverse spin of quarks, even within an unpolarized target. Coupling to the interference between the dipole and SM amplitudes, the transversity NEC produces characteristic $\sin \phi$ and $\cos \phi$ asymmetries in the energy pattern, providing clean access to the dipole couplings. Our combined analysis with both unpolarized and longitudinally polarized electron beams demonstrates that these observables can deliver precise and complementary constraints at the EIC.

This work initiates an exploration of energy-correlator observables in the search for new physics
(for an effort in other directions, see~\cite{Ricci:2022htc}). Compared with proposals based on transversity PDFs or dihadron production~\cite{Boughezal:2023ooo,Wen:2024cfu}, our NEC-based approach eliminates the need for nucleon polarization and relies entirely on inclusive calorimetric measurements, without particle identification or hadron reconstruction. These features simplify experimental implementation and reduce statistical uncertainties. They also enable the full utilization of high luminosity from the future EIC and LHeC as well as the use of existing HERA data. Furthermore, the resulting constraints may benefit from extensions to heavy-ion beams.

While the present study employs inclusive NECs, improved angular resolution could be achieved through track-based measurements of charged-hadron energy flux~\cite{Chen:2020vvp,Li:2021zcf,Jaarsma:2023ell,Lee:2023npz}, which would be valuable for precision studies on the dipole couplings. Furthermore, the chiral-odd NEC framework can be extended to polarized nucleons, with all such NECs sharing identical hard-scattering kernels, thus offering robust and complementary constraints. Beyond DIS, NECs could also be applied to processes such as Drell–Yan production at the LHC.
Detailed investigations along these lines are left for future work.

\begingroup
\renewcommand{\addcontentsline}[3]{}
\acknowledgments{{\it\noindent\color{blue} Acknowledgments} We thank Yu Jia, Yi Liao, Xiaohui Liu, Jianping Ma, Hongxi Xing for helpful discussions and comments.  X.B. Tong is supported by the Research Council of Finland, the Centre of Excellence in Quark Matter and projects 338263 and 359902, and by the European Research Council (ERC, grant agreements No. ERC-2023-101123801 GlueSatLight and No. ERC-2018-ADG-835105 YoctoLHC). H. L. Wang is supported
by the Grants No.\,NSFC-12035008. The content of this article does not reflect the official opinion of the European Union and responsibility for the information and views expressed therein lies entirely with the authors.}

{\it \noindent Note added} On the day this paper was submitted to arXiv, a preprint~\cite{Michel:2025afc} appeared in which the author proposed using a chiral-odd fracture function to probe Yukawa couplings at the LHC. In contrast to NECs, fracture functions typically involve measuring a specific type of hadron with momentum $\vec P_{h}$ in the TFR~(see e.g., \cite{Anselmino:2011ss,Chen:2023wsi,Chen:2024brp,Chen:2021vby,Chai:2019ykk}), which requires particle identification and tracking. While~\cite{Michel:2025afc} focused on the $\vec P_{h}$-integrated fracture functions,
we note that the  $\vec P_{h}$-dependent fracture functions (also referred to as extended fracture functions~\cite{Grazzini:1997ih}) are formally related to NECs through an inclusive energy sum rule recently established in \cite{Chen:2024bpj}.

\bibliographystyle{JHEP}
\bibliography{main.bib}
\makeatletter
\global\let\auto@bib\@empty
\global\let\auto@bib@innerbib\@empty
\makeatother
\endgroup

\clearpage
\title{Supplemental material for ``Nucleon energy correlators as a probe of light-quark dipole operators at the EIC''
}
\author{Yingsheng Huang}
\email{yingsheng.huang@outlook.com}
\affiliation{}

\author{Xuan-Bo Tong}\email{xutong@jyu.fi}
\affiliation{}

\author{Hao-Lin Wang}\email{whaolin@m.scnu.edu.cn}
\affiliation{}

\maketitle
\appendix
\onecolumngrid
\setcounter{secnumdepth}{2}
\allowdisplaybreaks

{ \tableofcontents}
\section{Quark transversity NEC}
In Eq.~(4) of our manuscript, we introduced the quark transversity NEC by definition. Here we provide additional details on its derivation and physical interpretation. We begin with the general quark NEC correlation matrix,
\begin{align}
  {\cal M}_{ij}(x,\vec v) &= \int \frac{d\eta^-}{2\pi} e^{-ix P^+ \eta^-} \langle P|\bar \psi_j(\eta^-){\cal E}(\vec v) \psi_i(0) |P\rangle~.
  \label{app:eq:NEEC_def}
\end{align}
where the three-vector  $\vec v=(\sin \theta \cos \phi, \sin \theta \sin \phi,\cos \theta)$ specifies the spatial direction of the measured energy flux. For brevity, we have suppressed the gauge links ${\cal L}_n$.

In analogy with the decomposition of collinear and TMD parton distributions~\cite{Jaffe:1996zw,Boussarie:2023izj}, hermiticity and parity constrain the NEC correlator to three independent Dirac structures at leading twist:
\begin{align}
  {\cal M}_{ij}(x,\vec v)=\frac{1}{2}\bigg[(\gamma^-)_{ij} {\cal M}^{[\gamma^+]}+(\gamma_5\gamma^-)_{ij} {\cal M}^{[\gamma^+\gamma_5]}-i \frac{(\sigma_{\alpha_\perp}^{~-} \gamma_5 )_{ij}}{2}{\cal M}^{[i \sigma^{\alpha_\perp+}\gamma_5 ]}\bigg]+{\cal O}\Big(\frac{M}{P^+}\Big)~,
  \label{eq:NECdecom}
\end{align}
where each component is obtained through the projection $\mathcal{M}^{[\Gamma]}=\frac{1}{2} \operatorname{Tr}[\Gamma \mathcal{M}]$. Because all leading-twist projections involve $\gamma^+$, they receive the dominant contributions in the high-energy limit ($P^+\rightarrow\infty$). Here, the chiral-even components ${\cal M}^{[\gamma^+]}$ and ${\cal M}^{[\gamma^+\gamma_5]}$, which describe the distributions of unpolarized or longitudinally polarized quarks, were discussed in detail in Ref.~\cite{Chen:2024bpj} for both unpolarized and polarized targets.

We focus on the chiral-odd component ${\cal M}^{[i \sigma^{\alpha_\perp+}\gamma_5 ]}$ , where $\Gamma^\alpha_\perp=i \sigma^{\alpha_\perp+}\gamma_5$ with $\alpha_\perp=1,2$ is the projector for quark transverse spin:
\begin{align}
  {\cal M}^{[i \sigma^{\alpha_\perp+}\gamma_5 ]}(x,\vec v)= \int \frac{d\eta^-}{2\pi} e^{-ix P^+ \eta^-} \langle P|\bar \psi(\eta^-) {\cal E}(\vec v) i \sigma^{\alpha_\perp+}\gamma_5  \psi(0) |P\rangle~.
\end{align}
The interpretation of quark transverse spin becomes explicit by rewriting
\begin{align}
  \Gamma^\alpha_\perp=\gamma^+ ({\cal Q}_{+}^\alpha-{\cal Q}_{-}^\alpha)~,
  \label{eq:projector}
\end{align}
where ${\cal Q}_{\pm}^\alpha=\frac{1}{2}\big(1\mp \gamma^5 \gamma^\alpha_\perp\big)$ projects onto quarks with transverse spin aligned or anti-aligned with the $\alpha_{\perp}$-direction~\cite{Jaffe:1996zw,Jaffe:1991ra}. Because transverse quark spin is invariant under longitudinal boosts, the corresponding NEC component ${\cal M}^{[i \sigma^{\alpha_\perp+}\gamma_5 ]}$ must transform covariantly under transverse rotations. For an unpolarized target, the only available transverse vector in the correlator is the transverse direction of the energy flow, ${\hat n}_{T}^\alpha\equiv\frac{{\bs v}_{\perp}^\alpha}{|{\bs v}_{\perp}|}=(\cos \phi,  \sin \phi)$, so the dependence on the transverse index $\alpha_{\perp}$ must appear through the unique pseudovector structure
\begin{align}
  \mathcal{M}^{\left[i \sigma^{\alpha_{\perp}+} \gamma_5\right]}(x,\vec v) = \epsilon_\perp^{\alpha\beta} \hat n_{T,\beta} h^t_1(x,\theta^2)~,
  \label{eq:def_sup}
\end{align}
where the transverse Levi-Civita tensor $\epsilon_{\perp}^{\alpha \beta}$ appears because the quark spin is a pseudovector. Eq.~\eqref{eq:def_sup} implies a $\sin\phi$ or $\cos\phi$ azimuthal   correlation between the quark's transverse polarization and the azimuthal direction of the energy flow.

The non-perturbative function $h_1^t(x, \theta^2)$, which we refer to as the quark transversity NEC, characterizes the strength of this azimuthal correlation as well as its dependence on $x$ and $\theta$. Using Eq.~(\ref{eq:projector}), it can be explicitly written as:
\begin{align}
  h^t_1(x,\theta^2)= \int \frac{d\eta^-}{4\pi} e^{-ix P^+ \eta^-} \langle P|\bar \psi(\eta^-) {\cal E}(\vec v) \gamma^+ \epsilon^\perp_{\alpha\beta} \hat n_{T\perp}^\beta ({\cal Q}_{+}^\alpha-{\cal Q}_{-}^\alpha)  \psi(0) |P\rangle~.
\end{align}
It describes the difference in probability of finding a quark whose transverse polarization is aligned with the pseudovector $ \epsilon^\perp_{\alpha\beta} \hat n_{T\perp}^\beta$ versus the opposite orientation. Here, the pseudovector $ \epsilon^\perp_{\alpha\beta} \hat n_{T\perp}^\beta$ is perpendicular to the transverse direction of the energy flow, $\hat{n}_{T}$. Thus, the quark transversity NEC $h_1^t(x,\theta^2)$ characterizes how the transverse polarization of a quark inside an unpolarized nucleon correlates with the direction of the energy flow detected in the target-fragmentation region.

\section{Factorization of the DIS energy pattern and the dipole-induced azimuthal modulations in the TFR}

\begin{figure}[htpb]
  \centering
  \includegraphics[width=0.6\textwidth]{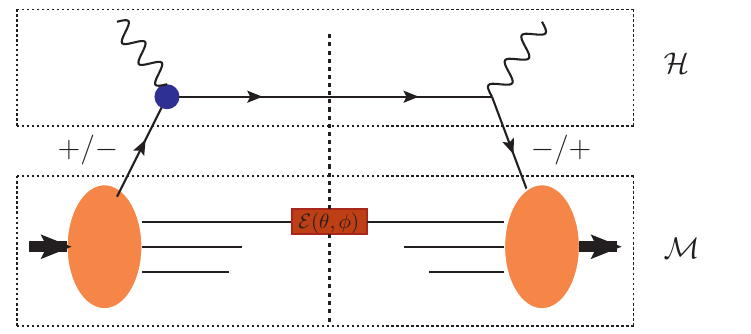}
  \caption{The factorized structure of the DIS energy pattern in the TFR with the inclusion of the light-quark dipole interactions. The blue blob denotes the dipole interactions. }
  \label{app:fig:NEC}
\end{figure}

In our manuscript, we show that the $\sin \phi$ and $\cos \phi$ modulations of the DIS energy pattern uniquely arise from the coupling of the quark transversity NEC to the dipole-SM interference, as presented in Eqs.~(6) and (10) of the main body. Here we provide additional discussion of the derivation from the factorization point of view. First, this derivation follows the general factorization theorem for the DIS energy pattern in the target fragmentation region (TFR) within the SM~\cite{Cao:2023oef,Chen:2024bpj}, with the additional inclusion of light-quark dipole interactions in the hard partonic scattering part $\mathcal{H}_{i j}$ and the quark transversity NEC in the nonperturbative parton correlation matrix $\mathcal{M}_{i j}$, as illustrated in Figure~\ref{app:fig:NEC}.

At the leading order of $\alpha_s$, the factorized structure takes the form:
\begin{align}
  &\frac{d \Sigma(\theta, \phi)}{d x_B d Q^2 }\propto
  \int_{x_B}^1 \frac{d x}{x}{L}_{\mu\nu}(l,l') {\cal H}_{ji}^{\mu\nu}\Big(\frac{x_B}{x}\Big) \delta(1-\frac{x_B}{x})
  {\cal M}_{ij}(x,\theta,\phi)~,
  \label{eq:factorization}
\end{align}
where $L_{\mu \nu}$ denotes the leptonic tensor. This formula captures the leading-power contributions in the TFR, valid in the kinematic regime $\theta P^{+} \ll Q$ under the Bjorken limit $Q \gg \Lambda_{\mathrm{QCD}}$.  In this region, the observed energy flow originates from the fragmentation of the target remnants after the virtual photon or
$Z$-boson knocks out a quark from the proton. The factorization arises because the long-distance target–fragmentation dynamics is well separated from the short-distance hard scattering, but remains strongly correlated with the struck quark. This long-distance correlation is encoded in the NEC correlator $\mathcal{M}_{i j}$, whereas the hard scattering between the quark and the virtual probe, $\mathcal{H}_{i j}$ is perturbatively calculable. Moreover, since we observe only the energy flow in the TFR and not in the current-fragmentation region, the transverse momentum of the struck quark is not resolved. As a result, the quark $k_{\perp}$-dependence in $\mathcal{M}_{i j}$ integrates out, yielding a collinear factorization structure. As demonstrated in Ref.~\cite{Chen:2024bpj}, the TFR factorization formula for the DIS energy pattern can be rigorously derived from the TFR factorization theorem for single-hadron inclusive DIS, originally proven by Collins in a seminal work~\cite{Collins:1997sr}.

In the SM, the hard scattering part ${\cal H}^{\mu\nu}_{ij}$ preserves chirality for light quarks. Therefore, the chiral-odd component $\mathcal{M}^{\left[i \sigma^\alpha \perp^{+} \gamma_5\right]}$ in Eq.~(\ref{eq:NECdecom}) does not contribute, and no azimuthal modulation arises for an unpolarized target. However, once the interference between the dipole operator and the SM amplitude is included, the hard part ${\cal H}^{\mu\nu}_{ij}$ contains an even number of $\gamma$-matrices, which allows a nonzero contraction with the chiral-odd structure:
\begin{align}
  \frac{d \Sigma(\theta, \phi)}{d x_B d Q^2 }\propto {L}_{\mu\nu}(l,l')\text{Tr}\Big[{\cal H}^{\mu\nu} i\sigma^{-\alpha_\perp} \gamma_5 \Big ]\epsilon_\perp^{\alpha\beta} \hat n_{T,\beta} h^t_1(x_B,\theta^2)~.
\end{align}
In our setup, the azimuthal angle $\phi$ of the energy flux is measured relative to the lepton scattering plane in the Breit frame. The transverse axes are chosen such that the leptons carry the transverse momentum $\bs l^\mu_\perp=-\bs l'^\mu_\perp=( Q\sqrt{1-y}/{y},0)
$. Because the partonic hard kernel ${\cal H}^{\mu\nu}$ does not depend on any transverse momentum, by rotational covariance, the contraction of the leptonic tensor with  ${\cal H}^{\mu\nu}$ must reduce to:
\begin{align}
  {L}_{\mu\nu}(l,l')\text{Tr}\Big[{\cal H}^{\mu\nu} i\sigma^{-\alpha_\perp} \gamma_5 \Big ]= {\cal C}_{1}  l^\alpha_\perp+{\cal C}_2  \epsilon_\perp^{\beta\alpha } l_{\perp,\beta}~.
\end{align}
where $\mathcal{C}_1$ and $\mathcal{C}_2$ are scalar functions of the invariants, linearly proportional to the real and imaginary parts of the quark dipole couplings, respectively.
Contracting this structure with the NEC correlator yields the characteristic  $\sin \phi$ and $\cos \phi$ modulations:
\begin{align}
  \frac{d \Sigma(\theta, \phi)}{d x_B d Q^2 }\propto \Big[{\cal C}_{1} \sin \phi +{\cal C}_2  \cos \phi\Big] h^t_1(x_B,\theta^2)~.
\end{align}
This demonstrates how the quark transversity NEC $h_1^t\left(x_B, \theta^2\right)$ provides the probe for the dipole-SM interference and generates the unique azimuthal modulations of the DIS energy pattern in the TFR.

\section{Evolution of the NEC moments in phenomenological analysis}

In this section, we describe how scale evolution is treated for NECs used in our phenomenological analysis.  The relevant momentum scale is identified as the virtuality $Q$ in the DIS process. As discussed in the main text, the observables considered in this work are expressed in terms of $\theta$-weighted moments of NECs. These moments can be related to $\boldsymbol{k}_\perp$-moments of TMDs, which allows us to connect NECs to existing phenomenological fits of TMDs.

We adopt the TMD fits as the initial conditions for the NEC moments at the scale $Q_0=2 \mathrm{GeV}$ and evolve them to other values of $Q$. This evolution is performed under the approximation that the $\theta$-moments of the NECs obey the same evolution equations as the corresponding NECs themselves.

We now justify this approximation by examining the evolution properties of the NECs. The unpolarized quark NEC $f_1\left(x, \theta^2\right)$ and the quark transversity NEC $h_1^t\left(x, \theta^2\right)$ obey the same renormalization group evolution equations as the unpolarized and transversity collinear quark PDFs, respectively. This follows from the fact that NECs can be formally expressed as collinear PDFs with insertions of energy-flow operators. Since the energy-flow operator acts on the inclusive final state, its insertion does not modify the ultraviolet structure of the operator and therefore does not affect the corresponding evolution equations~\cite{Chen:2024bpj}.

The observables considered in our analysis are related to $\theta$-moments of NECs. For quark transversity, the relevant moment is defined as
\begin{align}
  H_1^t(x) \equiv E_N \int d\theta^2\, |\sin\theta|\, h_1^t(x,\theta^2)
  = \int \frac{d^2\boldsymbol{k}_\perp}{2\pi}\,
  \frac{\boldsymbol{k}_\perp^2}{M}\,
  h_{1,\mathrm{TMD}}^{\perp}(x,\boldsymbol{k}_\perp^2).
\end{align}
which can be equivalently written as a $\bs k_\perp$-moment of the Boer–Mulders TMD $h_{1,\mathrm{TMD}}^{\perp}$. In principle, the evolution of this NEC moment differs from that of the NEC itself due to the appearance of additional inhomogeneous terms. From the known properties of the $\bs k_\perp$-moment of the BM TMD, the moment $H_1^t(x,Q)$ is related to the twist-3 quark--gluon correlator $T_{q,F}^{(\sigma)}(x,x,Q)$,
\begin{align}
  H_1^t(x,Q) \propto T_{q,F}^{(\sigma)}(x,x,Q),
\end{align}
whose QCD evolution has been derived in \cite{Kang:2012em}.

However, if one retains only the homogeneous term in the evolution equation, the moment $H_1^t(x,Q)$ evolves according to the same DGLAP equation as the transversity PDF~\cite{Artru:1989zv,Vogelsang:1997ak} and, correspondingly, the transversity NEC:
\begin{align}
  \frac{\partial}{\partial \ln Q^2} H_1^t(x,Q)
  = \frac{\alpha_s}{2\pi}
  \int_x^1 \frac{d\hat{x}}{\hat{x}}
  P_{q\to q}^{h_1}(\hat{x})\,
  H_1^t\!\left(\frac{x}{\hat{x}},Q\right),
\end{align}
with the splitting kernel
\begin{align}
  P_{q\to q}^{h_1}(\hat{x})
  = C_F\left[\frac{2\hat{x}}{(1-\hat{x})_+}
  + \frac{3}{2}\delta(1-\hat{x})\right].
\end{align}
In our phenomenological study, this homogeneous evolution is adopted as a reasonable approximation for estimating scale dependence. In other words, we assume the $\theta$-moments of the NECs follow the same evolution equations as the corresponding NECs themselves. Under this assumption, the evolution of the unpolarized NEC moment is automatically included through the collinear evolution encoded in the TMD fits.

\section{The complete results for energy-pattern cross section in the SM and SMEFT}
\label{app:strucfunc}

Here we summarize the complete results for the energy-pattern cross section in the SM and SMEFT. In this work, we consider the case where the target nucleon is unpolarized and the electron beam is unpolarized or longitudinally polarized. We are interested in the following components:
\begin{align}
  \Sigma(\theta,\phi)=& \Sigma_{UU}(\theta)+    \Sigma^{\sin \phi}_{UU}(\theta)\sin \phi +  \Sigma^{\cos \phi}_{UU}(\theta)\cos \phi \notag + \lambda_e \big[ \Sigma^{\sin \phi}_{LU}(\theta)\sin \phi +  \Sigma^{\cos \phi}_{LU}(\theta)\cos \phi \Big]~,
\end{align}
where the first and second subscripts denote the polarizations of the electron beam and the nucleon target, respectively. $\lambda
_e$ denotes the helicity of the electron beam. In the TFR, these energy-pattern cross sections can be factorized in terms of the associated NECs. The following results are given in the leading order of $\alpha_s$, and we have included all contributions from photon and $Z$ boson exchanges.

In the SM, only azimuthal-symmetric cross section exists and is yielded by the unpolarized quark NEC $f_1^q(x_B,\theta^2) $:
\begin{align}
  \nonumber
  &\frac{ \dd \Sigma_{UU}}{\dd x_B \dd Q^2}
  =\frac{2\pi\alpha^2_{\text{em}}}{ Q^4}\sum_{q} f_1^q(x_B,\theta^2)  \Bigg\{ Q_q^2 ( y^2-2y+2)  + \frac{2 Q^2}{Q^2+m_Z^2} \frac{Q_q}{(c_W s_W)^2}\Big[ g_A^e g_A^q(y-2) y- g_V^e g_V^q(y^2-2 y+2)\Big]
    \\
  &\qquad\qquad+ \frac{1}{(c_W s_W)^4} \left(\frac{Q^2}{Q^2+m_Z^2}\right)^2  \Big[ ( y^2-2y+2)  [(g_A^e)^2+(g_V^e)^2] [(g_A^q)^2+(g_V^q)^2] -4y(y-2)g_A^e g_A^q g_V^e g_V^q \Big]\Bigg\}~.
  \label{eq:UU_SM1}
\end{align}

\begin{figure}[ht!]
  \hspace{-0.5cm}
  \includegraphics[width=0.5\textwidth]{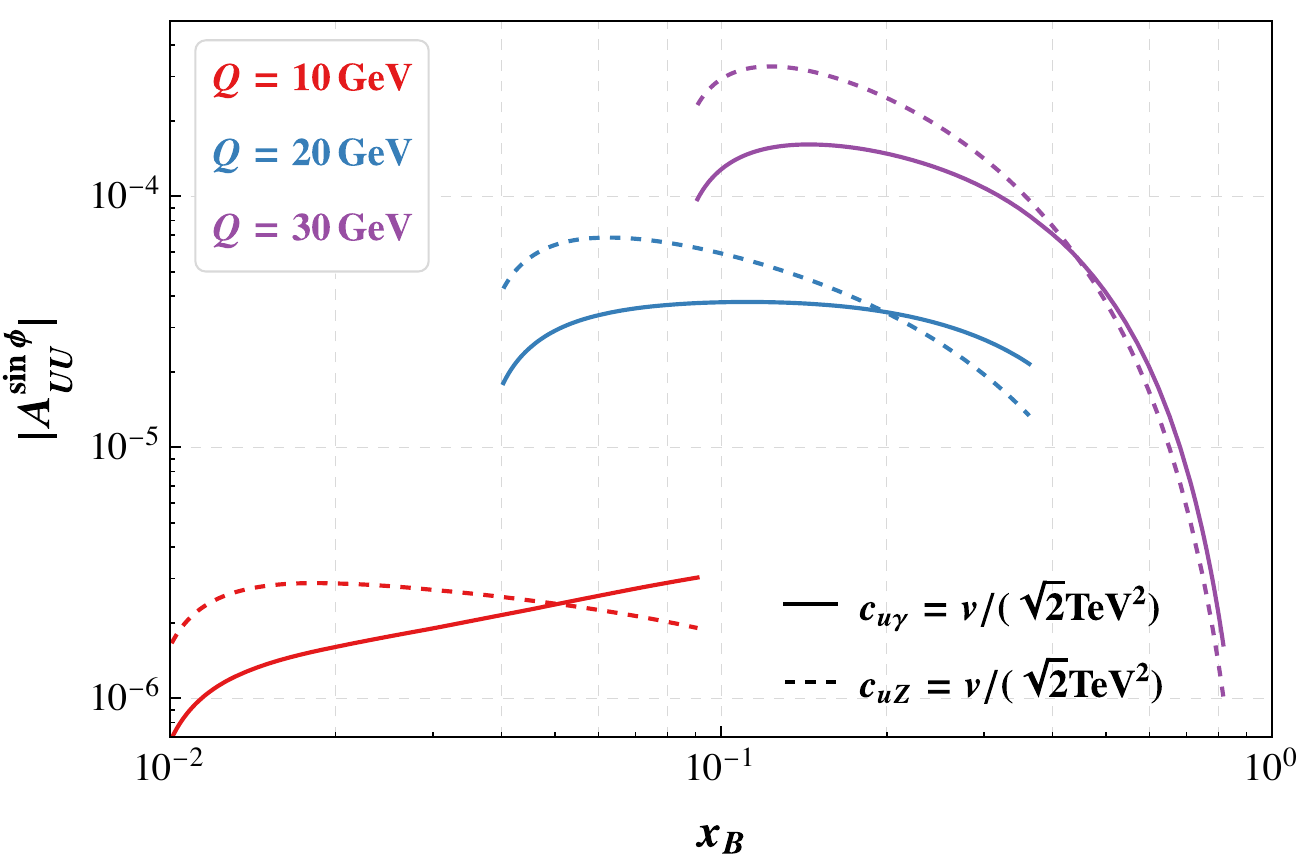}
  \includegraphics[width=0.5\textwidth]{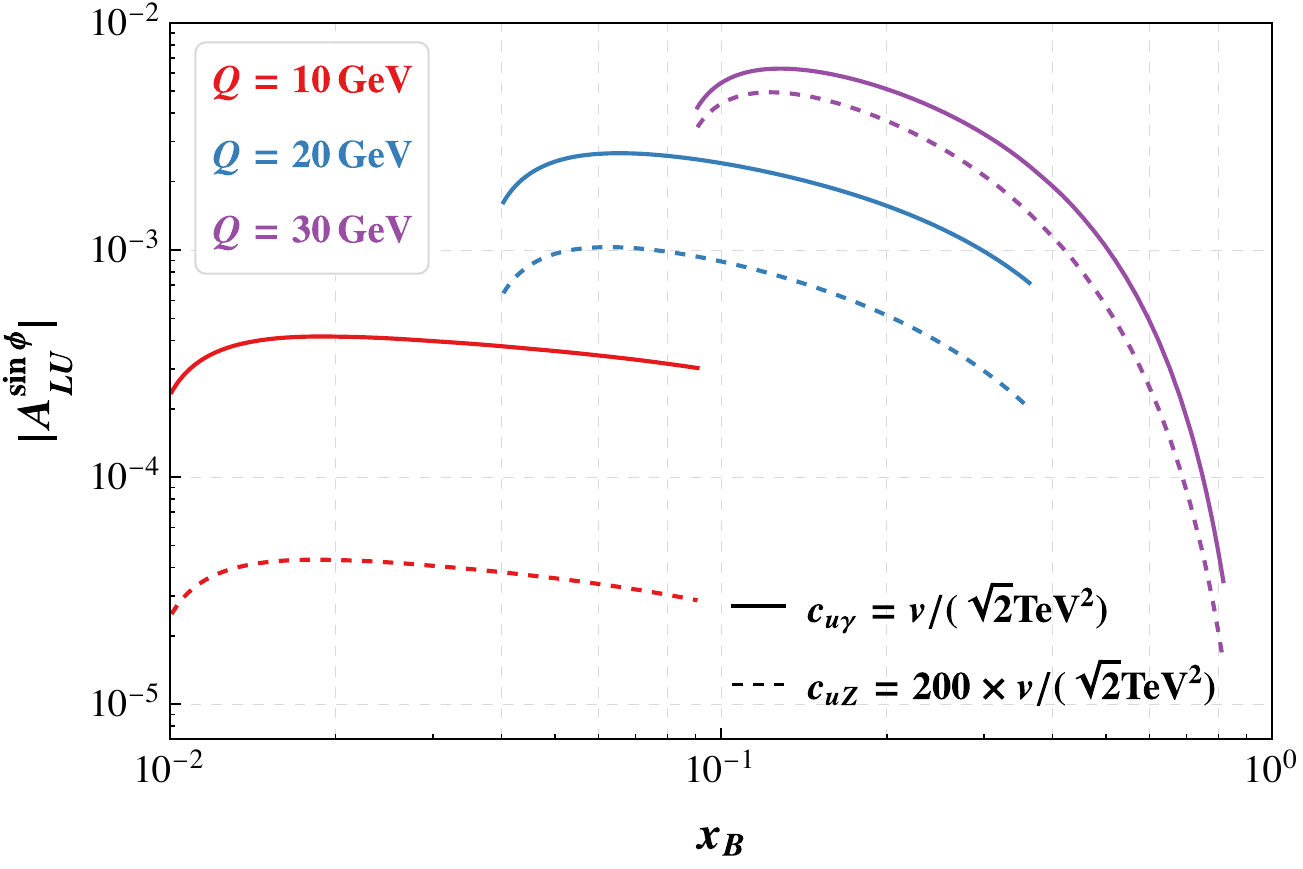}
  \caption{$|A_{UU}^{\sin \phi}|$ ($\textbf{Left}$) and $|A_{LU}^{\sin \phi}|$ ($\textbf{Right}$) as a function of the Bjorken variable $x_B$ for the up-quark dipole operators with different $Q$ at the EIC. The solid (dashed) lines correspond to setting $c_{u\gamma}\,(c_{uZ})=v/(\sqrt{2}\,{\rm TeV}^2)$. The center-of-mass energy is set to be $105\ \mathrm{GeV}$, and the inelasticity $y$ is limited to $[0.1,0.9]$.
  }
  \label{fig:asymmetry}
\end{figure}
In the SMEFT, the $\phi$-dependent cross sections with unpolarized and longitudinally polarized electrons are given by the dipole couplings and the quark transversity NEC $h_1^{t,q}(x_B, \theta^2)$.  For convenience, we present the results in mass basis:
\begin{equation}
  \label{eq:SMEFT:SSB}
  {\cal L} = c_{q\gamma}\left(\bar{q}_L \sigma_{\mu\nu} q_R\right) F^{\mu\nu} + c_{qZ}\left(\bar{q}_L \sigma_{\mu\nu} q_R\right) Z^{\mu\nu}+{\rm h.c.},
\end{equation}
where $q=u,d$ denote the up and down quarks. The dipole couplings $c_i$ are related to the dimensionless Wilson coefficients $C_{i}$ by
\begin{align}
  \begin{aligned}
    \label{eq:Crelations}
    c_{q\gamma} &= (v/{\sqrt{2}\Lambda^2})\left( c_W C_{qB} \pm s_W C_{qW} \right), \\
    c_{qZ} &= (v/{\sqrt{2}\Lambda^2})\left(-s_W C_{qB} \pm c_W C_{qW}  \right),
  \end{aligned}
\end{align}
where the plus (minus) sign is for the up (down) quark, and $v$ is the Higgs vacuum expectation value. The results are summarized as follows:
\begin{align}
  \nonumber
  \frac{ \dd \Sigma_{UU}^{\sin \phi}}{\dd x_B \dd Q^2}
  =&
  \frac{4\pi\alpha^2_{\text{em}}}{ ec_W s_W } \frac{y\sqrt{1-y} }{Q(Q^2+m_Z^2)}
  \sum_q h_1^{t,q}(x_B, \theta^2)
  \Bigg\{\Big[\frac{2-y}{y} g_A^{q}g_V^e+ g_V^{q}g_A^e\Big ]\frac{\Remy[c_{q\gamma}]}{c_W s_W}-Q_qg_A^e \Remy[c_{qZ}]
    \\
    &\qquad\qquad\qquad\qquad\qquad + \frac{1}{ (c_W s_W)^2 } \frac{Q^2}{Q^2+m_Z^2}
  \Big[ \frac{2-y}{y}[(g_A^e)^2+(g_V^e)^2] g_A^q + 2 g_A^e g_V^e g_V^q \Big] \Remy[c_{qZ}]\Bigg \},
  \\
  \nonumber
  \frac{ \dd\Sigma_{LU}^{\sin\phi}}{\dd x_B \dd Q^2}=&
  \frac{4\pi \alpha_{\text{em}}^2}{Q^3}\frac{y \sqrt{1-y}}{e} \sum_{q} h_1^{t,q}(x_B, \theta)
  \\
  \nonumber
  &\qquad\times\Bigg\{Q_q \Remy[c_{q\gamma}]- \frac{Q^2}{Q^2+m_Z^2}\frac{1}{c_W s_W}\bigg[\Big[\frac{2-y}{y} g_A^{q} g_A^e+ g_V^{q} g_V^e
    \Big ] \frac{\Remy[c_{q\gamma}]}{c_W s_W}-Q_qg_V^e \Remy[c_{qZ}]\bigg]
    \\
    &\qquad \qquad \qquad -\frac{1}{(c_W s_W)^3} \left(\frac{Q^2}{Q^2+m_Z^2}\right)^2
  \Big[ [(g_A^e)^2 + (g_V^e)^2] g_V^q + \frac{2(2-y)}{y} g_A^e g_A^q g_V^e  \Big] \Remy[c_{qZ}] \Bigg\},
  \\
  \nonumber
  \frac{ \dd \Sigma_{UU}^{\cos \phi}}{\dd x_B \dd Q^2}
  =&
  -\frac{4\pi\alpha^2_{\text{em}}}{ ec_W s_W } \frac{y\sqrt{1-y} }{Q(Q^2+m_Z^2)}
  \sum_q h_1^{t,q}(x_B, \theta^2)
  \Bigg\{\Big[\frac{2-y}{y} g_A^{q}g_V^e+ g_V^{q}g_A^e\Big ]\frac{\Immy[c_{q\gamma}]}{c_W s_W}-Q_qg_A^e \Immy[c_{qZ}]
    \\
    &\qquad\qquad\qquad\qquad\qquad + \frac{1}{ (c_W s_W)^2 } \frac{Q^2}{Q^2+m_Z^2}
  \Big[ \frac{2-y}{y}[(g_A^e)^2+(g_V^e)^2] g_A^q + 2 g_A^e g_V^e g_V^q \Big] \Immy[c_{qZ}]\Bigg \},
  \\
  \nonumber
  \frac{ \dd\Sigma_{LU}^{\cos\phi}}{\dd x_B \dd Q^2}=&
  -\frac{4\pi \alpha_{\text{em}}^2}{Q^3}\frac{y \sqrt{1-y}}{e} \sum_{q} h_1^{t,q}(x_B, \theta)
  \\
  \nonumber
  &\qquad\times\Bigg\{Q_q \Immy[c_{q\gamma}]- \frac{Q^2}{Q^2+m_Z^2}\frac{1}{c_W s_W}\bigg[\Big[\frac{2-y}{y} g_A^{q} g_A^e+ g_V^{q} g_V^e
    \Big ] \frac{\Immy[c_{q\gamma}]}{c_W s_W}-Q_qg_V^e \Immy[c_{qZ}]\bigg]
    \\
    &\qquad \qquad \qquad -\frac{1}{(c_W s_W)^3} \left(\frac{Q^2}{Q^2+m_Z^2}\right)^2
  \Big[ [(g_A^e)^2 + (g_V^e)^2] g_V^q + \frac{2(2-y)}{y} g_A^e g_A^q g_V^e  \Big] \Immy[c_{qZ}] \Bigg\}~.
\end{align}

Similar to the case of unpolarized electrons, we can introduce the azimuthal angle asymmetry for longitudinally polarized electrons by incorporating the spin asymmetry in the definition:
\begin{align}
  A_{LU}^{\sin \phi} =
  \frac{\pi}{2}\frac{\Sigma(\sin\phi >0)\vert_{\lambda_e=+1}  - \Sigma(\sin\phi<0)\vert_{\lambda_e=+1} - (\Sigma(\sin\phi>0)\vert_{\lambda_e=-1}-  \Sigma(\sin\phi<0)\vert_{\lambda_e=-1}) }{\Sigma(\sin\phi >0)\vert_{\lambda_e=+1}  +\Sigma(\sin\phi<0)\vert_{\lambda_e=+1} + (\Sigma(\sin\phi>0)\vert_{\lambda_e=-1}+  \Sigma(\sin\phi<0)\vert_{\lambda_e=-1})}~,
\end{align}
where $\Sigma(\sin\phi>0)$ represents the $\theta$-weighted energy pattern cross section integrated over the region with $\sin\phi>0$,
\begin{align}
  \Sigma(\sin\phi>0)\equiv\int ^{\theta_{\text{max}}^2}_{\theta_{\text{min}}^2}
  \dd \theta^2  |\sin \theta |\int^{2\pi}_0 \dd \phi\Sigma (\theta,\phi) \Theta(\sin\phi)~.
  \label{eq:weighting1}
\end{align}
The $A_{LU}^{\cos \phi}$ asymmetry can be defined in a similar manner.

\begin{figure}[t!]
  \begin{centering}
    \includegraphics[width=0.4 \textwidth]{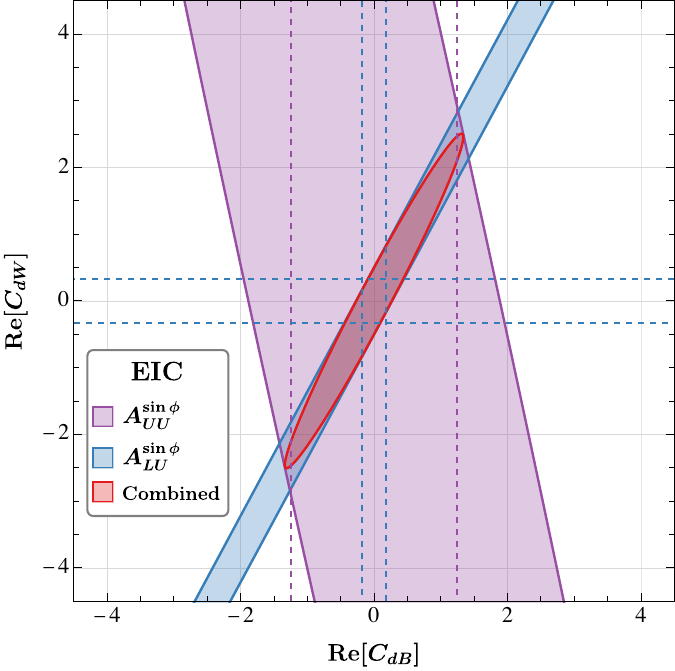}
    \caption{Projected $68\,\%$ C.L. constraints at the EIC on the two Wilson coefficients $\Remy[C_{dB}]$ and $\Remy[C_{dW}]$, assuming $\Lambda=1$ TeV. The limits are derived from the azimuthal asymmetries $A^{\sin\phi}_{UU}$ (purple), $A^{\sin\phi}_{LU}$ (blue), and their combination (red). The shaded regions represent the simultaneous two-coefficient constraints, while dashed lines show the limits for a single coefficient.
    }
    \label{fig:cons-CBW2}
  \end{centering}
\end{figure}

\begin{figure*}[b!!]
  \centering
  \includegraphics[width=0.4 \textwidth]{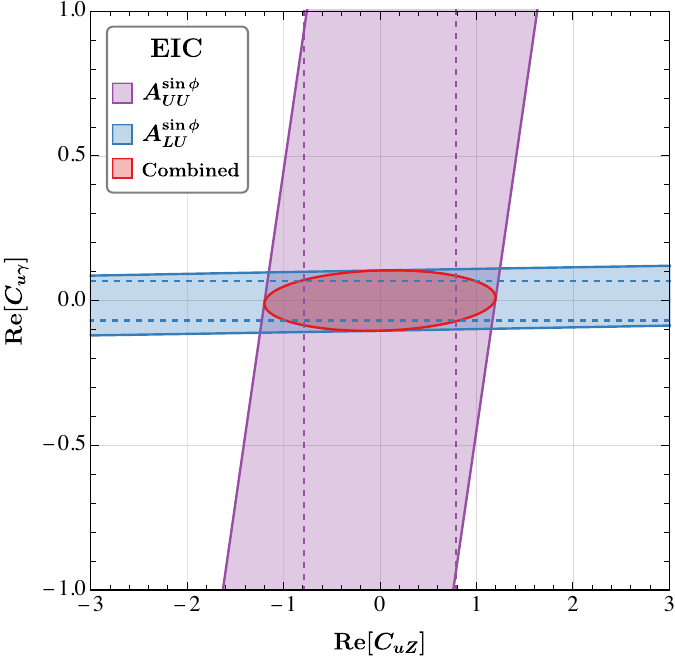}
  \qquad\qquad\quad
  \includegraphics[width=0.4 \textwidth]{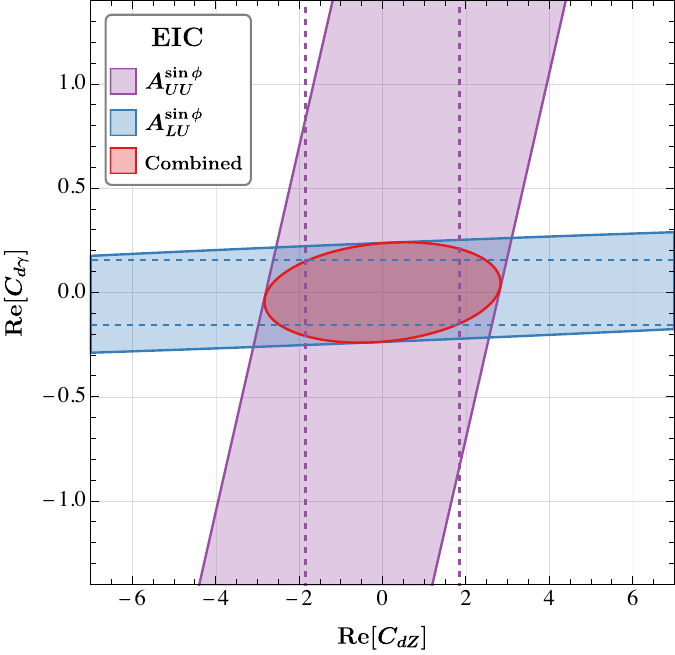}
  \caption{Similar to \cref{fig:cons-CBW2}, but for the dimensionless couplings $C_{q\gamma(Z)}\equiv c_{q\gamma(Z)}/(v/\sqrt{2} \textrm{TeV}^2)$.}
  \label{fig:cons-CZgamma}
\end{figure*}
For illustration, \cref{fig:asymmetry} shows the numerical results for $|A_{UU}^{\sin \phi}|$ and $|A_{LU}^{\sin \phi}|$ as a function of the Bjorken variable $x_B$ for the up-quark dipole operators with different $Q$ at the EIC kinematics.
We take $\sqrt{s} = 105\ \mathrm{GeV}$ to optimize luminosity and apply an inelasticity cut of $0.1 < y < 0.9$~\cite{AbdulKhalek:2021gbh}. The dimensionful dipole couplings are fixed to $c_{u\gamma} = v / (\sqrt{2}\,\mathrm{TeV}^2)$ for both asymmetries, with $c_{uZ} = v / (\sqrt{2}\,\mathrm{TeV}^2)$ for $|A_{UU}^{\sin \phi}|$ and $c_{uZ} = 200\,v / (\sqrt{2}\,\mathrm{TeV}^2)$ for $|A_{LU}^{\sin \phi}|$ to enhance visibility. We find $A_{UU}^{\sin \phi}>0$ for $c_{u\gamma}$ and $A_{UU}^{\sin \phi}<0$ for $c_{uZ}$, while $A_{LU}^{\sin \phi}<0$ for $c_{u\gamma}$ and $A_{LU}^{\sin \phi}>0$ for $c_{uZ}$.
The magnitude of both asymmetries grows with $Q$. In particular, $A_{LU}^{\sin \phi}$ from $c_{u\gamma}$ is free from the $Q^2/m_Z^2$ suppression, almost reaching $\mathcal{O}(10^{-2})$ at $Q=30$ GeV. Results for $d$-quark operators follow a similar pattern, with slightly reduced magnitudes due to the smaller $h_1^{t,d}$. In our analysis, we use TMDs as the input for the NECs and apply the parameterization from~\cite{Barone:2009hw}. The resulting asymmetries are consistent in magnitude with those obtained using the alternative fit in~\cite{Lu:2009ip}.

\section{Additional constraints on the dipole operators at the EIC}

In this section, we present the projected EIC sensitivities for $d$-quark dipole operators and for the dipole operators in the mass basis as given in Eq.~\eqref{eq:SMEFT:SSB}, providing complementary numerical results.

The constraints are derived from a $\chi^2$ analysis, with $\chi^2$ defined as
\begin{align}
  \chi^2 = \sum_{i} \left[\frac{A_{{\rm th},i}-A_{{\rm exp},i}}{\delta A_i}\right]^2,
\end{align}
where $A_{\text{th},i}$, $A_{\text{exp},i}$, and $\delta A_i$ denote the theoretical prediction, experimental measurement, and statistical uncertainty of the asymmetry in the $i$-th bin, respectively. For the EIC, with the cuts: $0.1\leq y \leq0.9$ and $0.01\leq x \leq 0.5$, we adopt different $Q$-binning schemes for $A_{UU}$ and $A_{LU}$ to account for their differing magnitudes.
For $A_{UU}$, a fixed bin width of $10$ GeV is used over the range $Q\in [10,60]$ GeV. In contrast, the $A_{LU}$ distribution is divided into $5$ GeV bins across $Q\in [10,50]$ GeV, with an additional bin spanning $Q\in [50,60]$ GeV.

\begin{table}[h!]
  \centering
  \setlength{\tabcolsep}{10pt}
  \begin{tabular}{l c c cccc}
    \hline\hline
    \textbf{Collider} & \textbf{$\bm{\sqrt{s}}$ [GeV]} & \textbf{$\bm{\mathcal{L}}$ [fb$^{-1}$]} & \textbf{$\bm{P_e}$} & \textbf{$\bm{Q}$ [GeV]} & \textbf{$\bm{y}$} & \textbf{$\bm{x_B}$} \\
    \hline
    EIC~\cite{AbdulKhalek:2021gbh}   & 105      & 100 & $70\%$ & $[10,60]$    & $[0.1,0.9]$ & $[0.01,0.5]$  \\
    HERA~\cite{Klein:2008di,H1:2015ubc}   & 318      & 0.4 & $40\%$ & $[30,150]$   & $[0.1,0.9]$ & $[0.01,0.5]$  \\
    LHeC~\cite{LHeC:2020van}   & 1300     & 50  & $80\%$ & $[100,1000]$ & $[0.1,0.9]$ & $[0.005,0.9]$ \\
    \hline\hline
  \end{tabular}
  \caption{A summary of the experimental parameters and kinematic cuts for the three colliders considered in this analysis. The parameters shown are the center-of-mass energy ($\sqrt{s}$), the integrated luminosity ($\mathcal{L}$), and the degree of electron beam polarization ($P_e$). The ranges for the photon virtuality ($Q$), inelasticity ($y$), and Bjorken variable ($x_B$) are also specified. }
  \label{tab:collider-params}
\end{table}
\begin{figure}[h!]
  \centering
  \includegraphics[width=0.8\textwidth]{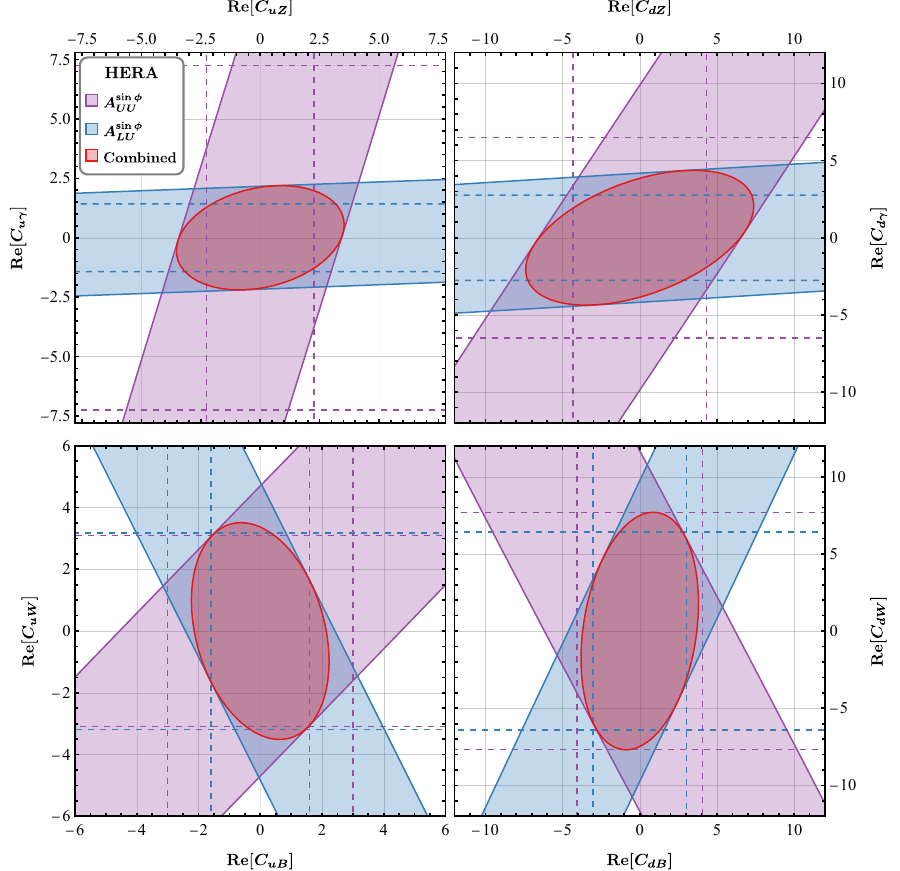}
  \caption{\raggedright The projected double-operator constraints at HERA with $68\%$ C.L. on the real parts of the couplings, distinguishing between the up-channel (left column) and down-channel (right column). The top row shows constraints in the $\text{Re}[C_{q\gamma}]$-$\text{Re}[C_{qZ}]$ planes ($C_{q\gamma(Z)}\equiv c_{q\gamma(Z)}/(v/\sqrt{2} \textrm{TeV}^2)$), while the bottom row shows the $\text{Re}[C_{qB}]$-$\text{Re}[C_{qW}]$ planes. All constraints are derived from the azimuthal asymmetries $A^{\sin\phi}_{UU}$ (purple) and $A^{\sin\phi}_{LU}$ (blue), assuming $\Lambda=1$ TeV. The red ellipses represent constraints from combining $A^{\sin\phi}_{UU}$ and $A^{\sin\phi}_{LU}$. The dashed purple and blue lines correspond to the single-operator constraints from $A^{\sin\phi}_{UU}$ and $A^{\sin\phi}_{LU}$ respectively.
  }
  \label{fig:HERA-cons}
\end{figure}
The projected constraints on ($\Remy[C_{dB}]$, $\Remy[C_{dW}]$) are presented in \cref{fig:cons-CBW2}, and those on ($\Remy[C_{qZ}]$, $\Remy[C_{q\gamma}]$) with $q=u,d$ are shown in \cref{fig:cons-CZgamma}. The latter couplings have been rescaled to dimensionless form via
\begin{align}
  C_{q\gamma(Z)}\equiv c_{q\gamma(Z)}/(v/\sqrt{2} \textrm{TeV}^2)~.
\end{align}The most stringent constraints from combining $A_{UU}^{\sin\phi}$ and $A_{LU}^{\sin\phi}$ are highlighted in red, reaching values in the range $\mathcal{O}(0.01)\sim\mathcal{O}(0.1)$ for $C_{q\gamma}$. Individual limits from single-operator (double-operator) analyses using $A_{UU}^{\sin\phi}$ and $A_{LU}^{\sin\phi}$ are indicated by purple and blue dashed lines (shaded regions), respectively. In some cases, the single-operator sensitivity from either $A_{UU}^{\sin\phi}$ or $A_{LU}^{\sin\phi}$ is too weak to be displayed; for instance, no limit is shown for $\Remy[C_{dW}]$ from $A_{UU}^{\sin\phi}$. The similarity between the ($\Remy[C_{dZ}]$, $\Remy[C_{d\gamma}]$)  and ($\Remy[C_{uZ}]$, $\Remy[C_{u\gamma}]$)  constraints arises from their identical analytic dependence on the asymmetry expressions, as shown in Sec.~\ref{app:strucfunc}. Due to the relatively smaller transversity NEC for $d$-quark, the combined constraints on operators involving $d$-quark are approximately a factor of two weaker than those for $u$-quark operators.

\begin{figure}[htbp]
  \centering
  \includegraphics[width=0.8\textwidth]{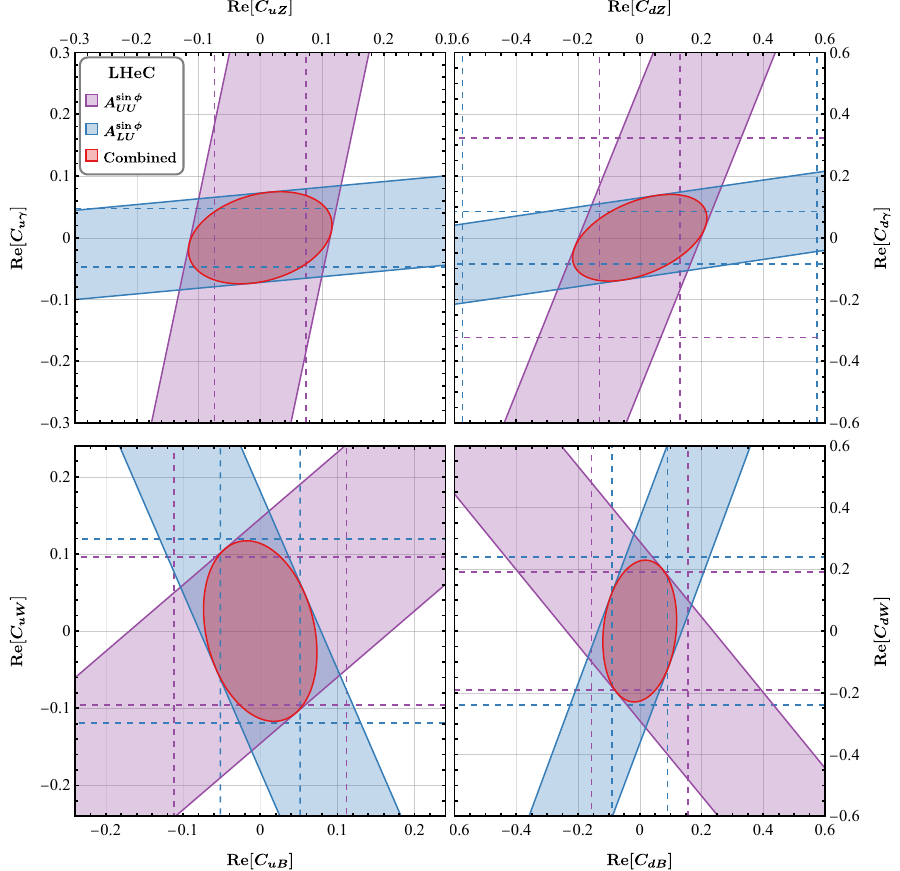}
  \caption{\raggedright The projected double-operator constraints at the LHeC with $68\%$ C.L. on the real parts of the couplings, distinguishing between the up-channel (left column) and down-channel (right column). The top row shows constraints in the $\text{Re}[C_{q\gamma}]$-$\text{Re}[C_{qZ}]$ planes ($C_{q\gamma(Z)}\equiv c_{q\gamma(Z)}/(v/\sqrt{2} \textrm{TeV}^2)$), while the bottom row shows the $\text{Re}[C_{qB}]$-$\text{Re}[C_{qW}]$ planes. All constraints are derived from the azimuthal asymmetries $A^{\sin\phi}_{UU}$ (purple) and $A^{\sin\phi}_{LU}$ (blue), assuming $\Lambda=1$ TeV. The red ellipses represent constraints from combining $A^{\sin\phi}_{UU}$ and $A^{\sin\phi}_{LU}$. The dashed purple and blue lines correspond to the single-operator constraints from $A^{\sin\phi}_{UU}$ and $A^{\sin\phi}_{LU}$ respectively.
  }
  \label{fig:LHeC-cons}
\end{figure}

\newpage

\section{Comparison of constraints among HERA, the EIC and the LHeC}
In addition to the EIC, this study analyzes the sensitivity of two other electron-proton colliders: HERA~\cite{Klein:2008di,Diaconu:2010zz} and the Large Hadron-Electron Collider (LHeC)~\cite{LHeC:2020van}. HERA is included since our setup does not require nucleon polarization, while the LHeC represents a future high-energy frontier with enhanced sensitivity to new physics. For our numerical analyses, we specifically consider the Run 6 phase of the LHeC. The parameters for these colliders are summarized in \cref{tab:collider-params}.

We now present the numerical results for HERA and the LHeC. Following the procedure used for the EIC, we derive constraints on the real parts of the relevant dimension-6 Wilson coefficients from a $\chi^2$ fit for the azimuthal asymmetries $A^{\sin\phi}_{UU}$ and $A^{\sin\phi}_{LU}$. This analysis is performed in $Q$ bins. For the LHeC, we use intervals of $100~\text{GeV}$ over the range $100 < Q < 1000~\text{GeV}$, while for HERA, intervals of $10~\text{GeV}$ are used over the range $30 < Q < 150~\text{GeV}$. The kinematic cuts applied in our analysis are also detailed in \cref{tab:collider-params}. The resulting projected constraints for HERA and the LHeC are presented in \cref{fig:HERA-cons} and \cref{fig:LHeC-cons}, respectively. These figures detail one-dimensional limits on individual operators and two-dimensional constraints on operator pairs involving both up and down quarks. These sensitivities are derived from fits using the $A_{UU}^{\sin\phi}$ and $A_{LU}^{\sin\phi}$ asymmetries individually, as well as from a combined fit. All constraints are presented in both the electroweak basis (e.g., $C_{uB}, C_{uW}$) and the mass basis (e.g., $C_{u\gamma}, C_{uZ}$).

\begin{figure}[h]
  \begin{centering}
    \includegraphics[width=0.8\textwidth]{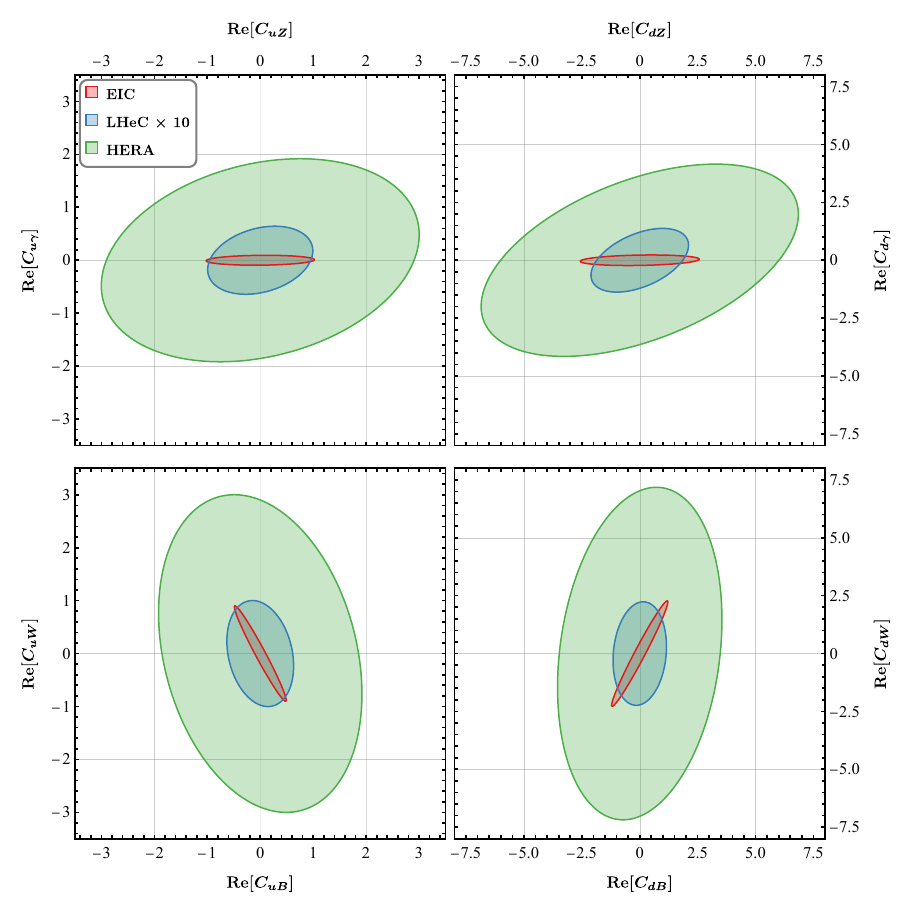}
    \caption{Comparison of projected (EIC, LHeC) and existing (HERA) constraints at $68\%$ C.L. on the real parts of the couplings, shown for the EIC (red), LHeC (blue, scaled by a factor of $10$), and HERA (green). The layout distinguishes between the up-channel (left column) and down-channel (right column). The top row corresponds to the $\text{Re}[C_{q\gamma}]$-$\text{Re}[C_{qZ}]$ planes ($C_{q\gamma(Z)}\equiv c_{q\gamma(Z)}/(v/\sqrt{2} \textrm{TeV}^2)$) and the bottom row corresponds to the $\text{Re}[C_{qB}]$-$\text{Re}[C_{qW}]$ planes. Each contour represents a combined constraint derived from the azimuthal asymmetries $A^{\sin\phi}_{UU}$ and $A^{\sin\phi}_{LU}$, assuming $\Lambda=1$ TeV.
    }
    \label{fig:collider-comparison}
  \end{centering}
\end{figure}

\cref{fig:collider-comparison} presents the final two-operator constraints from the EIC (red), HERA (green), and LHeC (blue) for a direct comparison. For visualization purposes, the LHeC ellipses are scaled by a factor of $10$. The figure clearly illustrates the superior reach of the LHeC, a consequence of both its high center-of-mass energy and its large integrated luminosity. In contrast, HERA's potential is limited by its much smaller luminosity, though it still achieves a constraining power slightly weaker than that of the EIC thanks to its own high center-of-mass energy. Ultimately, both LHeC and HERA benefit from their access to a higher $Q^2$ range, resulting in similarly shaped constraining ellipses, whereas the EIC's limited $Q^2$ range induces stronger correlations between the coefficients.

\newpage

\section{Discussions on the uncertainties from non-perturbative inputs}

In this section, we evaluate the impact of theoretical uncertainties associated with the nonperturbative inputs on our projected constraints. The transversity NEC, $h_1^t$, which serves as the crucial input for our asymmetry calculation, is estimated with the first transverse moment of the Boer-Mulders quark TMD function, $\mathscr{h}_1^{\perp q}(x, \bs k_\perp^2)$. Consequently, uncertainties in the extraction of the Boer-Mulders TMD propagate to our final constraints on the dipole couplings. For comparison and simplicity, we adopt the scale evolution implemented in the TMD fits for the corresponding NEC moments in this section.

To assess the robustness of our results, we consider two factors affecting the nonperturbative uncertainties: (1) the uncertainties within the Barone \textit{et al.}~fit~\cite{Barone:2009hw} we adopted in our analysis, and (2) the impact of using different Boer-Mulders TMD fits.

\subsection{Uncertainties from the Barone \textit{et al.} Fit}

\begin{figure}[h]
  \centering
  \includegraphics[scale=0.84]{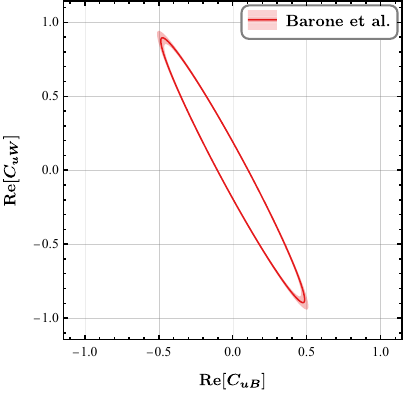}
  \label{fig:TMD_para_comp}
  \includegraphics[scale=0.86]{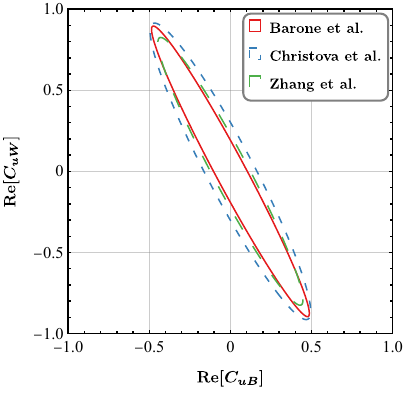}
  \caption{Left: Constraints based on the fit of Barone \emph{et al.}, including the associated uncertainties. Right: Comparison of the constraints on the dipole couplings obtained using different fits of the Boer–Mulders TMD as input (Here, we have used the central values).
  }
  \label{fig:uncertainty}
\end{figure}

In our primary analysis, we utilized the parameterization of the Boer-Mulders TMD provided by Barone \textit{et al.}~\cite{Barone:2009hw}. This fit is based on a combined analysis of COMPASS and HERMES data and relies on the ansatz that the Boer-Mulders function is proportional to the Sivers function, $f_{1T}^{\perp q}$:
\begin{equation}
  \mathscr{h}_1^{\perp q}(x, \bs k_\perp^2) = \lambda_q \mathscr{f}_{1 T}^{\perp q}(x, \bs k_\perp^2).
\end{equation}
The dominant uncertainty in this extraction resides in the flavor-dependent normalization factors, $\lambda_q$. The best-fit values and their associated uncertainties are given by:
\begin{equation}
  \lambda_u = 2.0 \pm 0.1, \quad \lambda_d = -1.111 \pm 0.001.
\end{equation}
To estimate the impact of these uncertainties on our projected constraints, we generate an envelope with the upper and lower values of $\lambda_q$. We demonstrate the effects of this envelope on the projected constraints in Fig.~\ref{fig:uncertainty}~(Left). The variation in the TMD parameters leads to only a slight broadening of the allowed region, indicating that the uncertainties in this fit do not significantly affect the competitiveness of our projected bounds.
\subsection{Impact of different phenomenological fits}

To assess the impact of different phenomenological  fits, we compared the constraints derived using the Barone \emph{et al.} fit with those obtained from two other independent parameterizations of Boer-Mulders TMD available in the literature:
\begin{itemize}
  \item \textbf{Christova \textit{et al.}~\cite{Christova:2020ahe}:} A more recent extraction based on COMPASS SIDIS data. Notably, this analysis does not assume the proportionality ansatz between the Boer–Mulders and Sivers functions used in Barone \emph{et al.}.
  \item \textbf{Zhang \textit{et al.}~\cite{Zhang:2008ez}:} An extraction utilizing data from unpolarized $p+D$ Drell-Yan processes, providing a complementary constraint from a different scattering channel.
\end{itemize}
For this comparison, we adopted the central best-fit values from each parameterization. The resulting constraints are displayed in Fig.~\ref{fig:uncertainty}~(Right).

We observe consistent bounds derived from all three parameterizations, indicating our results are not sensitive to the choice of nonperturbative inputs. This insensitivity can be understood as follows. The asymmetry is sensitive primarily to the first transverse moment of the Boer–Mulders function, weighted by $\bs k_\perp^2/M_N^2$. While the detailed functional forms of $\mathscr{h}_1^{\perp q}(x, \bs k_\perp^2)$ differ among the fits, they predict similar effective transverse widths. Furthermore, our analysis involves an integration over a wide range of Bjorken-$x$, which tends to wash out local differences in the $x$-dependence of the various parametrizations.

The above analysis demonstrates that the projected sensitivity to the light-quark dipole couplings is stable against the current theoretical uncertainties associated with the nonperturbative Boer–Mulders TMD inputs.

\section{Comparison with other collider probes on light-quark dipole operators}
In this section, we provide a detailed comparison of our projected sensitivities with existing or proposed constraints from other collider probes. We focus specifically on the constraints derived  from existing Drell-Yan data at the LHC~\cite{Boughezal:2021tih} and from a proposed dihadron  measurement at the EIC~\cite{Wen:2024cfu}, as these represent the most relevant competing methods for probing light-quark dipole moments at colliders.

\subsection{Comparison with  Drell-Yan constraints at the LHC}

We first compare our projections with constraints derived from the Drell-Yan process ($pp \to Z/\gamma^*(\to\ell^+\ell^-)+X$) at the LHC~\cite{Boughezal:2021tih}. As discussed in our manuscript, in the Drell-Yan process, the contribution from dimension-6 dipole operators appears only at $\mathcal{O}(\Lambda^{-4})$ in the cross-section, due to the chiral structure of the interaction. This leads to two limitations. First, the signal is suppressed relative to our observable, which arises from the $\mathcal{O}(\Lambda^{-2})$ interference between the dipole operator and the SM amplitude. This suppression, however, can be compensated by the higher energy of the LHC, compared to the EIC and HERA in our proposal. Second, the observables in Ref.~\cite{Boughezal:2021tih} suffer from contamination by various other SMEFT operator contributions. Specifically, non-dipole dimension-6 operators can contribute at $\mathcal{O}(\Lambda^{-2})$ and significantly dominate over the dipole terms, while dimension-8 operators also enter at $\mathcal{O}(\Lambda^{-4})$. Consequently, deriving robust constraints from Drell-Yan requires the assumption that these contaminating operators are absent. In contrast, our observable is uniquely sensitive to the dipole–SM interference at the leading $\mathcal{O}(\Lambda^{-2})$ order. This feature not only enhances sensitivity but also protects the analysis from contamination by other potential UV effects.
Furthermore, because the Drell-Yan cross-section depends on the squared modulus of the couplings ($|C|^2$), it cannot resolve the complex phase of the dipole operators. Our method allows for the separate extraction of the real and imaginary parts via $\cos\phi$- and $\sin\phi$-weighted asymmetries.

\begin{figure}[h]
  \centering
  \includegraphics[scale=0.7]{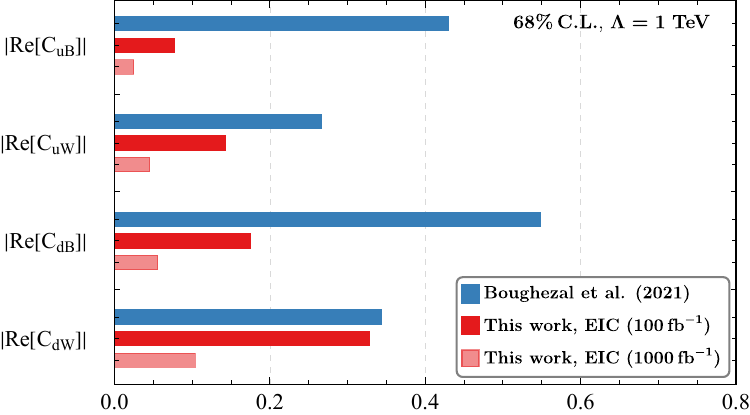}
  \caption{Comparison between our projected single-operator constraints at the EIC and those reported in Ref.~\cite{Boughezal:2021tih} based on Drell–Yan measurements at the LHC. Here, we have included both the $100\ \textrm{fb}^{-1}$ and $1000\ \mathrm{fb}^{-1}$ EIC configurations for comparison.}
  \label{fig:compare_DY}
\end{figure}
In Fig.~\ref{fig:compare_DY}, we show the quantitative comparison. We include the reported bounds from Ref.~\cite{Boughezal:2021tih} (rescaled to $68\%$ C.L.) alongside our projections for both the conservative ($\mathcal{L} = 100~\mathrm{fb}^{-1}$) and high-luminosity ($\mathcal{L} = 1000~\mathrm{fb}^{-1}$) EIC configurations. Even with the conservative luminosity, our observable yields constraints stronger than the current LHC Drell-Yan bounds. With the high-luminosity configuration, the sensitivity is further improved, providing a powerful and theoretically clean probe of the dipole couplings.

\subsection{Comparison with the dihadron approach for the EIC}

Ref.~\cite{Wen:2024cfu} (Wen \textit{et al.}) achieves a nonzero dipole-SM interference by observing collinear dihadron production in the DIS with an unpolarized proton beam ($e p \to e (h_1 h_2) X$).

A direct comparison between the results in Ref.~\cite{Wen:2024cfu} and our main manuscript is not straightforward due to differences in the assumed luminosity ($\mathcal{L}=1000\ \mathrm{fb}^{-1}$ in Ref.~\cite{Wen:2024cfu} vs. $\mathcal{L}=100\ \mathrm{fb}^{-1}$ in our baseline) and operator conventions. To facilitate a like-for-like comparison, we have performed a new analysis of our observable assuming the high-luminosity configuration ($\mathcal{L}=1000\ \mathrm{fb}^{-1}$) and converted the results of Ref.~\cite{Wen:2024cfu} into our operator convention using the relation:
\begin{align}
  C_{q\gamma} = \frac{e}{\sqrt{2}} \left( \frac{1\,\mathrm{TeV}}{v} \right)^2 \Gamma^q_\gamma~,
\end{align}
where $v \approx 246$ GeV is the Higgs vacuum expectation value, and $ \Gamma^q_\gamma$ is the dipole coupling in the convention of Ref.~\cite{Wen:2024cfu}.

\begin{figure}[h]
  \centering
  \includegraphics[scale=0.65]{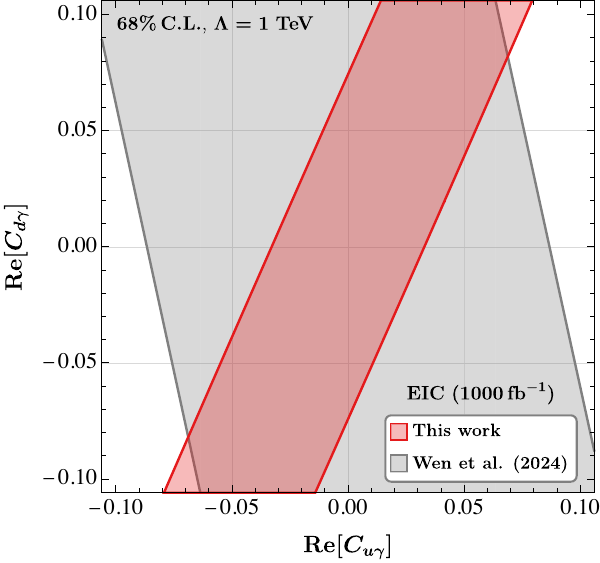}
  \caption{A comparison between the projected constraints reported in Ref.~\cite{Wen:2024cfu} based on dihadron production at the EIC and our corresponding results. Here, we have adopted the same integrated luminosity, ${\cal L}=1000~\mathrm{fb}^{-1}$, for a meaningful comparison.}
  \label{fig:compare_Wen}
\end{figure}
In Fig.~\ref{fig:compare_Wen}, we compare the combined constraints on $\mathrm{Re}[C_{u\gamma}]$ and $\mathrm{Re}[C_{d\gamma}]$. As shown in the figure, our observable yields a significantly smaller allowed region, excluding a substantial portion of the parameter space allowed by the dihadron analysis. While both our observables and Ref.~[18] exhibit strong correlations between up- and down-quark dipole couplings, they yield different flat directions. Consequently, a combined analysis could substantially mitigate these correlations. Additionally, under a single-operator assumption, Ref.~\cite{Wen:2024cfu} estimates the magnitude of the constraint on $C_{q\gamma}$ (after conversion to our convention) to be approximately $0.1$. Our projected sensitivities for the same luminosity improve upon this estimation by approximately one order of magnitude.

Finally, we note that the observable proposed in this work offers distinct experimental advantages. It requires neither the polarized proton beams utilized in Ref.~\cite{Boughezal:2023ooo} nor the semi-inclusive measurements involving particle identification and multi-particle tracking required for the dihadron method in Ref.~\cite{Wen:2024cfu}.

\end{document}